\documentclass[twocolumn,trackchanges,astrosymb]{aastex62}

\received{\today}
\revised{}
\accepted{}
\submitjournal{PASP}

\shorttitle{The quantum efficiency and internal diffraction behavior of Si:As mid-IR detectors}
\shortauthors{G\'asp\'ar et al.}

\begin{document}

\title{The quantum efficiency and diffractive image artifacts of Si:As IBC mid-IR detector arrays at 5 $-$ 10 $\mu$m: Implications for the {\it JWST}/MIRI detectors}

\correspondingauthor{Andr\'as G\'asp\'ar}
\email{agaspar@arizona.edu}

\author[0000-0001-8612-3236]{Andr\'as G\'asp\'ar}
\author[0000-0003-2303-6519]{George H.\ Rieke}
\affil{Steward Observatory and the Department of Astronomy, The University of Arizona, 933 N Cherry Ave, Tucson, AZ, 85750, USA}

\author[0000-0002-2421-1350]{Pierre Guillard}
\affil{Sorbonne Universit\'e, CNRS, UMR 7095, Institut d'Astrophysique de Paris, 98bis bd Arago, 75014 Paris, France
}
\affil{Institut Universitaire de France, Minist\'ere de l'Education Nationale, de l'Enseignement Sup\'erieur et de la Recherche, 
1 rue Descartes, 75231 Paris Cedex 05, France}

\author[0000-0003-0589-5969]{Daniel Dicken}
\author{Ren\'e Gastaud}
\affil{AIM, CEA, CNRS, Universit\'e Paris-Saclay, Universit\'e Paris Diderot, Sorbonne Paris Cit\'e, F-91191 Gif-sur-Yvette, France}

\author[0000-0002-8909-8782]{Stacey Alberts}
\author[0000-0002-8456-6188]{Jane Morrison}
\affil{Steward Observatory and the Department of Astronomy, The University of Arizona, 933 N Cherry Ave, Tucson, AZ, 85750, USA}

\author[0000-0001-5644-8830]{Michael E.\ Ressler}
\affil{Jet Propulsion Laboratory, California Institute of Technology, 4800 Oak Grove Drive, Pasadena, CA 91109, USA}

\author[0000-0003-3149-7980]{Ioannis Argyriou}
\affil{Institute of Astronomy, Celestijnenlaan 200D - box 2401, 3001 Leuven, Belgium}

\author{Alistair Glasse}
\affil{UK Astronomy Technology Centre, Royal Observatory, Edinburgh, Blackford Hill, Edinburgh EH9 3HJ, UK}

\begin{abstract}
Arsenic doped back illuminated blocked impurity band (BIBIB) silicon detectors
have advanced near and mid-IR astronomy for over thirty years; they have high
quantum efficiency (QE), especially at wavelengths longer than 10 $\micron$, and a 
large spectral range. Their radiation hardness is also an asset for space based 
instruments. Three examples of 
Si:As BIBIB arrays are  used in the Mid-InfraRed Instrument (MIRI) of the James Webb Space 
Telescope ({\it JWST}), observing between 5 and 28 $\micron$. In this paper, we 
analyze the parameters leading to high quantum efficiency (up to $\sim$ 60\%) 
for the MIRI devices between 5 and 10 $\mu$m. We also model the cross-shaped artifact that 
was first noticed in the 5.7 and 7.8 $\micron$ {\it Spitzer}/IRAC images and has since also 
been imaged at shorter wavelength ($\le 10~\micron$) laboratory tests of the MIRI detectors. 
The artifact is a result of internal reflective diffraction off the pixel-defining metallic contacts to the readout detector circuit. The low absorption in the arrays at the shorter wavelengths 
enables photons diffracted to wide angles to cross the detectors and substrates multiple times. 
This is related to similar behavior in other back illuminated solid-state detectors with poor 
absorption, such as conventional CCDs operating near 1 $\mu$m.  We investigate the properties 
of the artifact and its dependence on the detector architecture with a quantum-electrodynamic 
(QED) model of the probabilities of various photon paths. Knowledge of the artifact properties 
will be especially important for observations with the MIRI LRS and MRS spectroscopic modes.
\end{abstract}


\section{Introduction}
\label{sec:intro}

Developed in the 1980’s \citep[e.g.,][]{petroff80, petroff84, petroff85, stetson86}, arsenic 
doped extrinsic silicon (Si:As) Blocked-Impurity-Band (BIB - or Impurity Band Conductor - IBC) 
detector arrays are widely used in the field of mid-infrared astronomy. Multiple space 
missions have made them their detector of choice due to particular attributes: their wide 
spectral region coverage (5-28 $\mu$m), high quantum efficiency, low dark current, stable performance 
for an extended  period of time, low heat dissipation (which is particularly important for 
cryo-cooled space missions), and nuclear radiation hardness. The sizes and performance of 
arrays of these detectors have increased over time. The Short Wavelength Spectrometer on 
the Infrared Satellite Observatory (ISO), launched in 1995,  utilized a 1 $\times$ 12 array 
\citep{leech2003}, the MSX satellite, launched in 1996, had  8 $\times$ 192 arrays 
\citep{mill1994}, while IRS and MIPS on the Spitzer Space Telescope had 128 $\times$ 128 devices 
\citep{vancleve1995} and the IRAC/Spitzer 5.7 and 7.8 $\mu$m \citep{hora04} and the Akari 
\citep{onaka07} arrays were 256 $\times$ 256. The latest generation detectors are 1024 
$\times$ 1024, such as the 12 and 23 $\mu$m arrays of  the WISE mission \citep{mainzer08} 
and the three focal plane arrays (FPAs) \citep{ressler08} mounted in the Mid-InfraRed 
Instrument (MIRI) of the James Webb Space Telescope ({\it JWST}).

Much of the theory on the operation of these detectors was published in the 1980's, shortly 
after they were invented, i.e.: \citet{petroff84, petroff85, stetson86, szmulowicz1987, szmulowicz1988}. 
Since then there has been substantial progress in the detector quality, primarily through 
improved control of minority impurities, resulting in successful fabrication of devices with 
much thicker infrared-active layers than previously \citep{love04}. The original efforts are 
largely successful in describing the properties of these improved devices at the longer 
wavelengths. However, testing of detectors with these thicker layers showed that the behavior 
of responsivity vs. bias voltage requires introduction of a diffusion length of $\sim$ 2.5 
$\mu$m (rather than the negligible diffusion length in the original theory), or the response 
at low bias and wavelengths near the peak of response would fall short of measurements 
\citep{rieke15}.  This change is required because at wavelengths $\gtrsim$ 15 $\mu$m, most of 
the photons are absorbed in the first $\sim$ 10 microns of the infrared-active layer. At low 
bias, the field does not penetrate across this layer and the photo-electrons must diffuse across 
a field-free zone to be collected and produce a signal. 

In addition, the detectors have been utilized down to wavelengths as short as 5 $\mu$m, far 
from their peak response at 12 - 25 $\mu$m. 
Quantum Efficiencies (QE)\footnote{QE is defined as the fraction or percentage of incident photons 
converted to electrons} of $>$ 40\% are expected in the 5 - 8 $\mu$m range \citep{love05,woods11}. 
Such applications complement InSb and HgCdTe photodiodes that until recently have had high 
performance only at wavelengths short of 5 $\mu$m.  Most notable of these applications was 
the IRAC/Spitzer channel 3 and 4 detector arrays \citep{fazio04, hora04}, developed by 
Raytheon \citep{estrada98}. Pre-launch measurements indicated that the IRAC detectors reached 
a QE of 37\% (following revisions after launch) and 56\% at 5.6 and 8.0 $\mu$m, respectively 
\citep{pipher04}. However, on orbit 
the achieved performance fell short of expectations from these values; the equivalent 
instrument throughputs were only 45\% and 61\% of the  pre-launch predictions, respectively. 
This discrepancy was traced to a substantial portion of the signal appearing over an extended 
region on the detector. This signal was included in the laboratory measurements that integrated 
over a large fraction of the array, but was excluded in the on-orbit measurements, which were analyzed 
with  point-source photometry \citep{hora04}. This behavior raises a question about whether the 
expected QE was actually achieved.

On-orbit images with IRAC showed that this extended response appeared as an image artifact, 
which the IRAC publications termed ``banding'' \citep{hora04,pipher04}.  The artifact can be 
described as a cross along the pixel columns and rows centered on the sources. Various tests 
were executed on a sister detector of the IRAC flight hardware at the University of Rochester 
to find the origin of the artifact \citep{pipher04}. External optical effects were ruled out 
by the placement of a dark mask around the laboratory pinhole, which did not mitigate the 
``banding.'' Their tests also showed that the strength of the artifact increases with decreasing 
wavelength and that light is scattered not just into the bands, but also into the entire detector
area. These results indicate that the artifact is not due to an issue with the readout electronics 
but is internal to the detector. The IRAC Instrument handbook \citep{iracihb} describes the banding as a result 
of diffraction off the rectilinear grid of conductive pads and multiple scattering within the detector.
This conclusion indicates that the behavior is similar to scattering observed in the near infrared 
with conventional CCDs \citep{gull2003,ryon19}; i.e., that it is a general issue with back-illuminated 
solid state detectors being operated in spectral regimes where they have poor absorption. 

This paper complements the evaluation of the long wavelength behavior of Si:As IBC detectors described 
in \citet{rieke15}. We will discuss both the short wavelength QE and the imaging artifacts. We start 
with a description of the detector architecture (Section 2). We then derive the wavelength dependence 
of the quantum efficiency and show good agreement with the behavior of the flight detectors for MIRI 
on {\it JWST}, including quantum efficiencies of up to $\sim$ 60\% in the 5 $-$ 10 $\mu$m range for arrays 
with suitable anti-reflection coatings (Section 3). The remainder of the paper focuses on the cause 
and morphology of the spreading of images in the shorter wavelengths, i.e., the ``banding,'' or the more 
descriptive term ``the cross artifact'' (Section 4). The work and modeling in this paper will inform 
the on-orbit image analysis for MIRI, which will in turn be used to refine the models to enhance the 
calibration of this instrument, particularly of its spectrometers. 

\section{Detector architecture}
\label{sec:architecture}

\begin{figure}
\begin{center}
\includegraphics[width=0.47\textwidth]{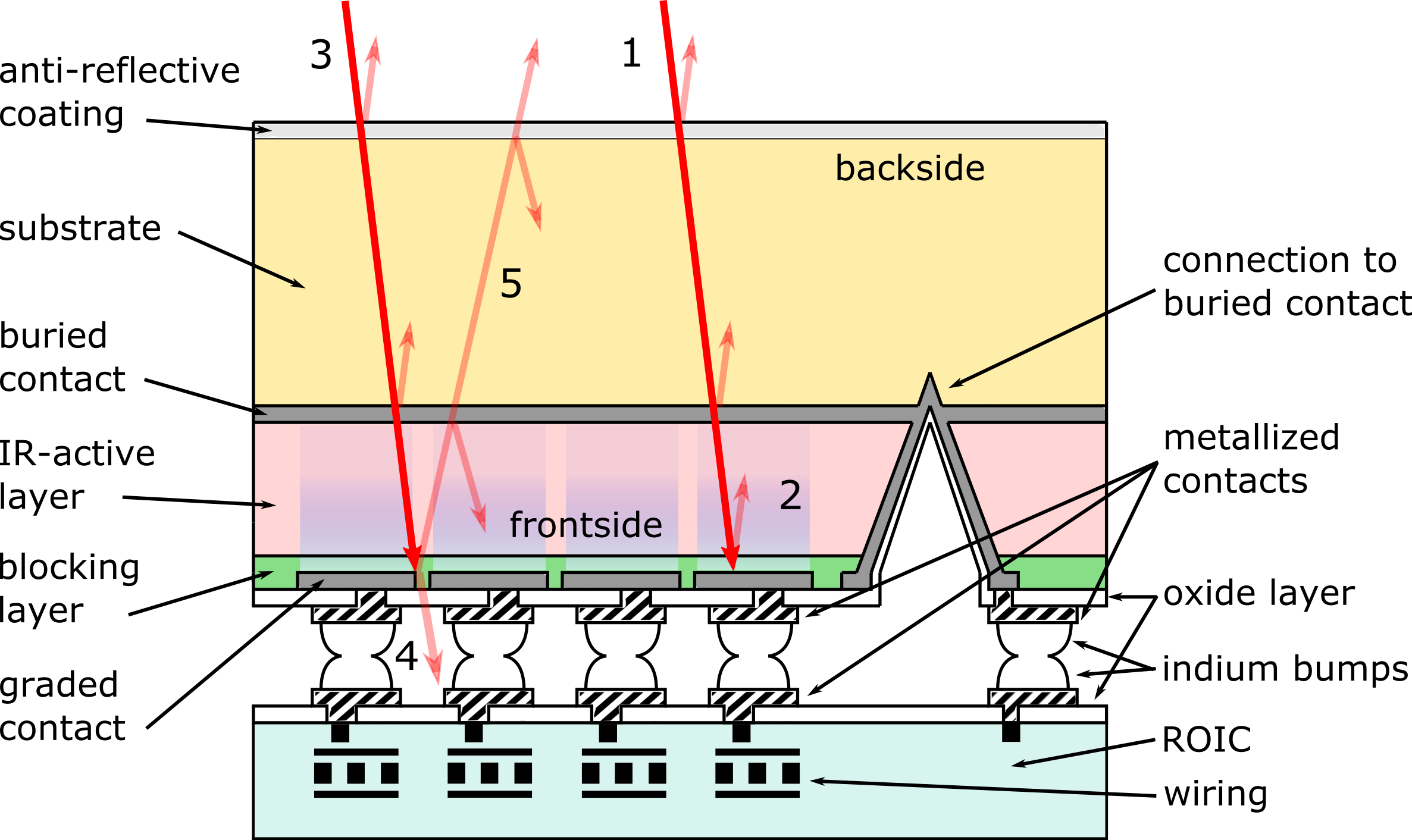}
\caption{This drawing (not to scale) shows the general architecture of the MIRI Si:As arrays. Four 
pixels are shown with contacts and indium bumps to allow connection to their amplifiers in the 
readout integrated circuit (ROIC). The detector bias is established between the output contacts 
and the buried one.  The individual electric fields of the four pixels are shown with a transparent blue 
gradient. The detectors are illuminated through their ``backside'', i.e. through the substrate that 
supports them rather than directly into the detectors, i.e. the ``frontside.'' Two possible photon 
paths are also shown. Path (1) is the traditional fate of photons: they undergo reflective losses 
at the entrance and reflective and absorptive losses at the buried contact. If they are then absorbed 
in the IR-active layer, they create a photoelectron that traverses this layer and the blocking layer 
to be collected at a contact.  Those that are not absorbed in the IR-active layer are reflected off 
the metallized contacts (2) and traverse the detector layers in reverse. Photons along path (3) have 
a similar trajectory but they are incident on a gap between the metallized contacts and either pass 
into the gap between detector and ROIC wafers (4) or are diffracted to wide angles (5), some of 
which are sufficiently off-axis to lead to total internal reflection and trapping within the detector 
wafer. These wide-angle photons may be detected far from the original input point. 
\label{fig:arch}}
\end{center}
\end{figure}

In Figure~\ref{fig:arch}, we show the general architecture of modern Si:As IBC detector arrays. 
The detectors are grown on a substrate and illuminated through it. They consist of (1) a buried 
contact to establish an electrical field, (2) an IR-active layer that is heavily doped with arsenic 
to create an impurity band where the photons are absorbed, elevating free electrons into the conduction 
band, (3) a high purity layer that blocks dark current from the impurity band but allows passage of
photoelectrons in the conduction band, and (4) an output contact. For the MIRI baseline arrays, the 
substrate is 420 $\mu$m thick, the IR-active layer is 35 $\mu$m, and the blocking layer is 
$\sim$ 4 $\mu$m. The substrate is of a high resistivity (i.e., low impurity content) silicon wafer. 
The arsenic impurity (a donor) in the IR-active layer has a concentration of $7 \times 10^{17}$ 
cm$^{-3}$, while acceptor-type impurities are at a level of about $1.5 \times  10^{12}$ cm$^{-3}$. 
This low level of the minority impurity is critical to allow the electric field (shown in blue
shading in the figure) to penetrate the IR-active layer for complete collection of photoelectrons 
at a bias level that does not trigger avalanche gain (and the resulting increase in noise). The 
$\sim$ 4 $\mu$m thick blocking layer just before the output contacts is of high purity and hence 
has no impurity band. As a result, it blocks thermal currents, while allowing free transit to 
photoelectrons, which have been elevated into the conduction band (which, of course, is continuous 
from the IR-active layer through the blocking one).  When the photoelectrons reach the contacts, 
they are sensed by integrating amplifiers in the readout integrated circuit (ROIC), connected 
through indium bump bonding hybridization (i.e., cold welding of indium bumps deposited on the detector and the readout wafers). 

Figure~\ref{fig:arch} also illustrates some fates for incident photons. There are reflective losses 
at the entrance to the substrate and both reflective and absorptive ones in the buried contact. Near 
the peak of response (12 - 24 $\mu$m) close to all the photons that survive through these layers are 
absorbed in the IR-active layer, either in the first pass (downward in Figure~\ref{fig:arch}) or 
after reflection from a metallized contact. However, a significant fraction of photons ($\ge 10\%$) 
that are not absorbed in the initial downward pass are lost within an inter-pixel gap, while
others undergo diffraction due to the pixel lattice structure (the fraction of photons that diffract
depends on the detector architecture and the wavelength of the photons). 
In the region of peak response, very few of these photons will get through the IR-active layer on 
their upward trajectory and are therefore most likely detected in the same pixel as they entered. 
However, if the IR-active layer has low absorption many of them will escape back upward into the 
substrate. Many of those will be sufficiently off-normal to undergo total internal reflection when they 
encounter the upper edge of the substrate, leading to their being trapped in the silicon. These
photons are responsible for the large scale of the imaging artifact.

\section{Response and quantum efficiency}
\label{sec:response}

Si:As IBC detectors can offer systems level advantages through their ability to provide good performance 
from 5 through $\sim$ 27 $\mu$m. To quantify this performance, in this section we consider the QE and 
response in the 5 $-$ 10 $\mu$m range, where the arsenic absorption coefficient is less than the values 
at the peak of response from 12 $-$ 20 $\mu$m. The low values of the absorption cross section between 
5 and 10 $\mu$m have led to a presumption that the QE of the detectors would be low there. However, QEs 
of up to $\sim$ 60\% are achieved by the MIRI detectors; our model explores the parameters that make such high 
levels possible. 

\subsection{Modeling the response}

\noindent
Our model of the short wavelength response of Si:As IBC detectors uses the parameters of the MIRI 
``baseline'' devices as described above and in more detail by 
\citet{love04,love05,love2006,ressler08,rieke15,ressler2015}.  We assume that the wavelength-dependent 
response is determined through a combination of (1) the absorption characteristics of the Si:As, 
(2) the anti-reflection coating on the entrance surface, (3) the optical characteristics of the 
buried contact, and (4) geometric factors.

\begin{figure}
\begin{center}
\includegraphics[width=0.47\textwidth]{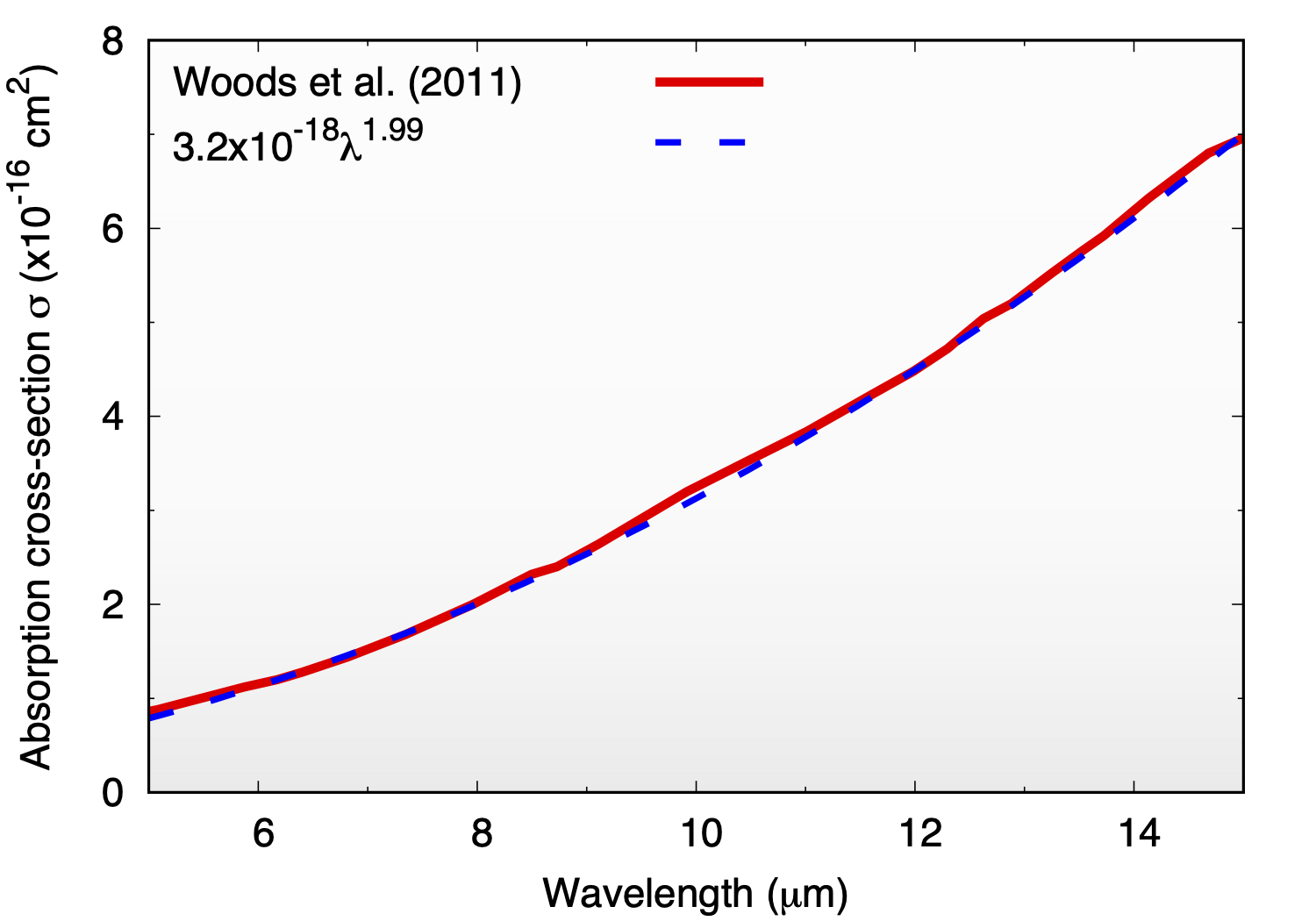}
\caption{The absorption cross section for Si:As from \citet{woods11}, along with a simple fit.
\label{fig:xsec}}
\end{center}
\end{figure}

There are a number of determinations of the absorption cross section of As in Si 
\citep{petroff85, geist89, woods11}.  \citet{geist89} shows that taking the cross section to be 
0.7 times the formula provided by \citet{petroff85} provides an excellent fit to the data between 
500 and 750 cm$^{-1}$ (13.3 -- 20 $\mu$m). The \citet{petroff85} fit follows a dependence of $\propto \lambda^{1.85}$ between 
5 and 25 $\mu$m. However, \citet{geist89} finds that the slope in the theoretical study of \citet{coon86} 
is steeper, as $\propto \lambda^{2.25}$. The measurements of \citet{woods11} are the only ones extending 
down to 5 $\mu$m (and below). They agree with those of \citet{geist89} and \citet{petroff85} in most of the region 
of overlap. They are 
inaccurate beyond 25 $\mu$m, failing to reflect the steep drop in sensitivity that begins at 
$\sim$ 27 $\mu$m. However, the region from 5 -- 15 $\mu$m is the range of interest for this paper. We show 
their results for these wavelengths in Figure~\ref{fig:xsec} along with a fit proportional to $\lambda^{1.99}$. 

\citet{ilaiwi1990} show that the shape of the theoretically predicted absorption coefficient depends 
critically on the assumed potential describing the screening behavior of the impurities. However, 
their calculations agree very well with each other for the isotropic hydrogenic, Yukawa, and 
H\"ulth\'en potentials, all of which also agree with the $\propto \lambda^{1.85 - 1.99} $ behavior. 

We consider the absorption coefficient to be well determined, therefore; to be specific, we have used 
the values from \citet{woods11}. The errors on these values are estimated at 6\% at 10 $\mu$m but much 
larger at 2 $\mu$m; we have assumed between 5 and 10 $\mu$m they scale linearly, so they are 
12\% at the former wavelength. 

The MIRI arrays have one of two single-layer AR coatings of ZnS, in one case optimized for 6 $\mu$m and in the 
other for 16 $\mu$m. We have 
modeled the array where this coating is optimized for 16 $\mu$m, since the high absorption coefficient 
around this wavelength allows us to determine the effects of the buried contact unambiguously. We have 
taken cryogenic refractive indices for silicon from \citet{li1980} and for ZnS from \citet{Hawkins2004}. 
The predicted  wavelength-dependent quantum efficiency shows two peaks, i.e., one at 16.3 $\mu$m 
corresponding to $\lambda/4$ and another at 5.8 $\mu$m for $3 \lambda/4$ (the wavelengths do not differ 
by exactly a factor of three because of the wavelength dependence of the refractive indices). These 
values are a result of tuning the modeled  thickness of the AR coating to provide the best match to 
the observed peaks, deriving a value of 1.95 $\mu$m (in good agreement with the measured value). 

\begin{figure}
\begin{center}
\includegraphics[width=0.47\textwidth]{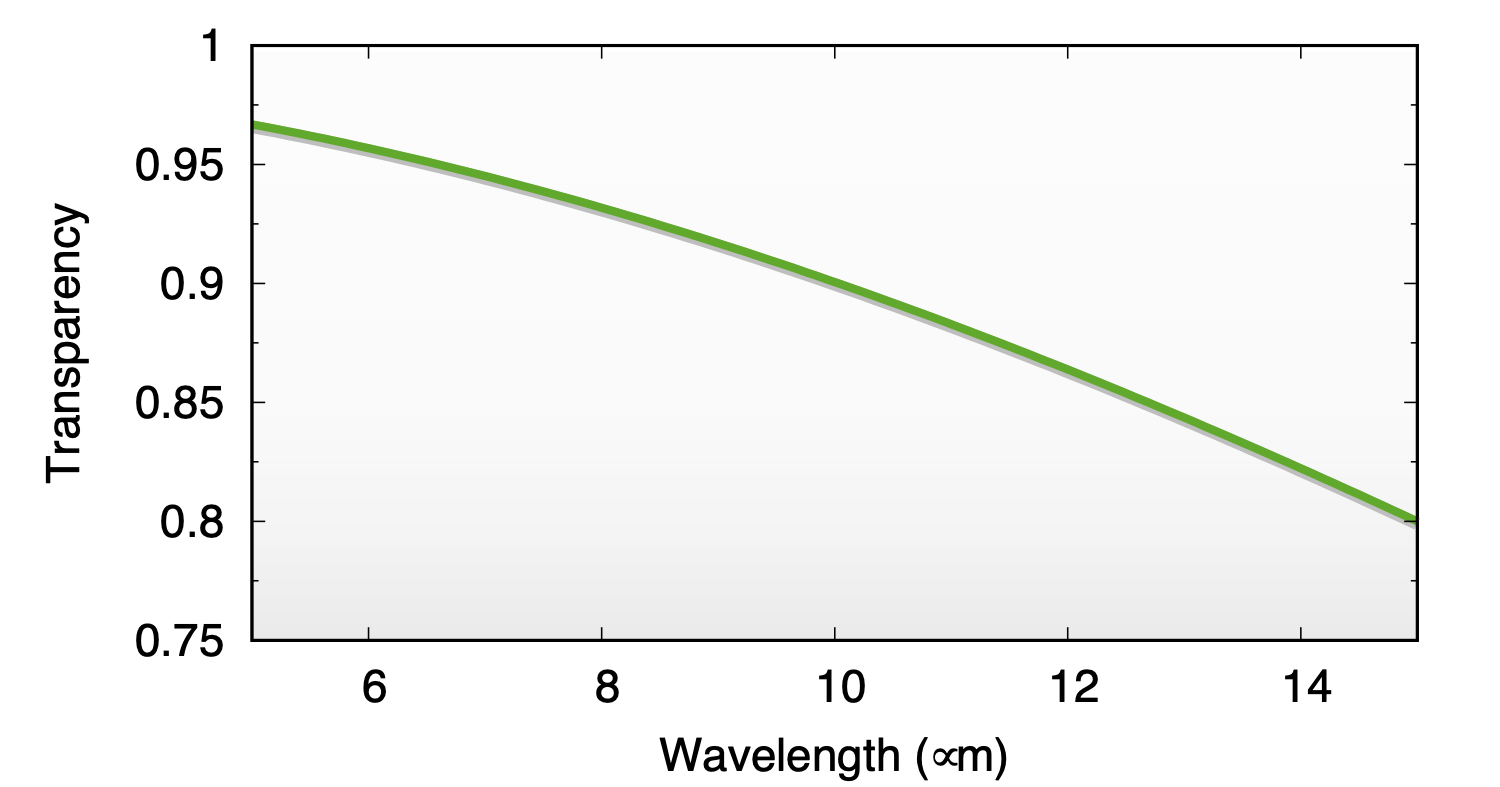}
\caption{The absorptance of the buried contact. The plot shows that the buried contact is much more 
absorptive at the longer wavelengths. The higher transparency of the contact below 10 $\mu$m helps 
explain the relatively high QE at these wavelengths. The high transparency is also important for our 
modeling in Section~\ref{tracerpaths}, as it determines the fraction of photons that are able to travel 
through the buried contact multiple times and hence be absorbed in the IR-active layer 
and detected far from their entry point into the array.
\label{fig:trans}}
\end{center}
\end{figure}

From early-on, it was recognized that the buried ``transparent'' contact was not fully transparent; 
\citet{petroff85} estimated that it was 25\% absorptive for one of their back-illuminated detectors. 
These contacts consist of a layer doped with a shallow impurity, such as arsenic, antimony, or phosphorus, so 
they show a decrease in absorption toward short wavelengths, just as photoconductive detectors based 
on such impurities do. The optical behavior of such contacts has been modeled by \citet{hoelzlein80} 
but is seldom included in modeling detectors. These models show that they both reflect and absorb 
incoming energy. We have obtained the absorptive component for typical contacts from M. G. Stapelbroek 
(private communication, see Figure~\ref{fig:trans}). The fringing model described by \citet{Argyriou2020} 
indicates reflection at the level of $\sim$ 4\% from $\sim$ 5 to $\sim$ 9 $\mu$m (beyond 9 $\mu$m, the 
detector absorption dominates the optical behavior). We have assumed 4\% reflection at all wavelengths 
for the QE modeling (we ignore reflectance for the diffraction pattern modeling in the later part of the paper). 

The primary geometric property affecting the detector quantum efficiency is the inter-pixel gaps in the 
backside contacts. The pixels are 25 $\mu$m square and have pixel-to-pixel gaps of $\sim$ 2 $\mu$m. The 
model described below assumes that all photoelectrons freed in the infrared-active layer are collected at a contact 
because of the fringing effect on electric fields for closely spaced electrodes \citep[e.g.][]{lisowski2009}. 
However, photons that penetrate to the contacts are reflected back through the detector layer, where 
additional absorption adds to the photo-current. In our model, the reflected signal is decreased to match 
the areal coverage of the contacts. A small additional absorption occurs for photons reflected back from 
the buried contact or detector backside when they pass through the IR-active layer. The model will include a
total of two complete passes through the detector (i.e.\ four passes through the IR-active layer).

\begin{figure}
\center
\includegraphics[angle=00,width=3.5in]{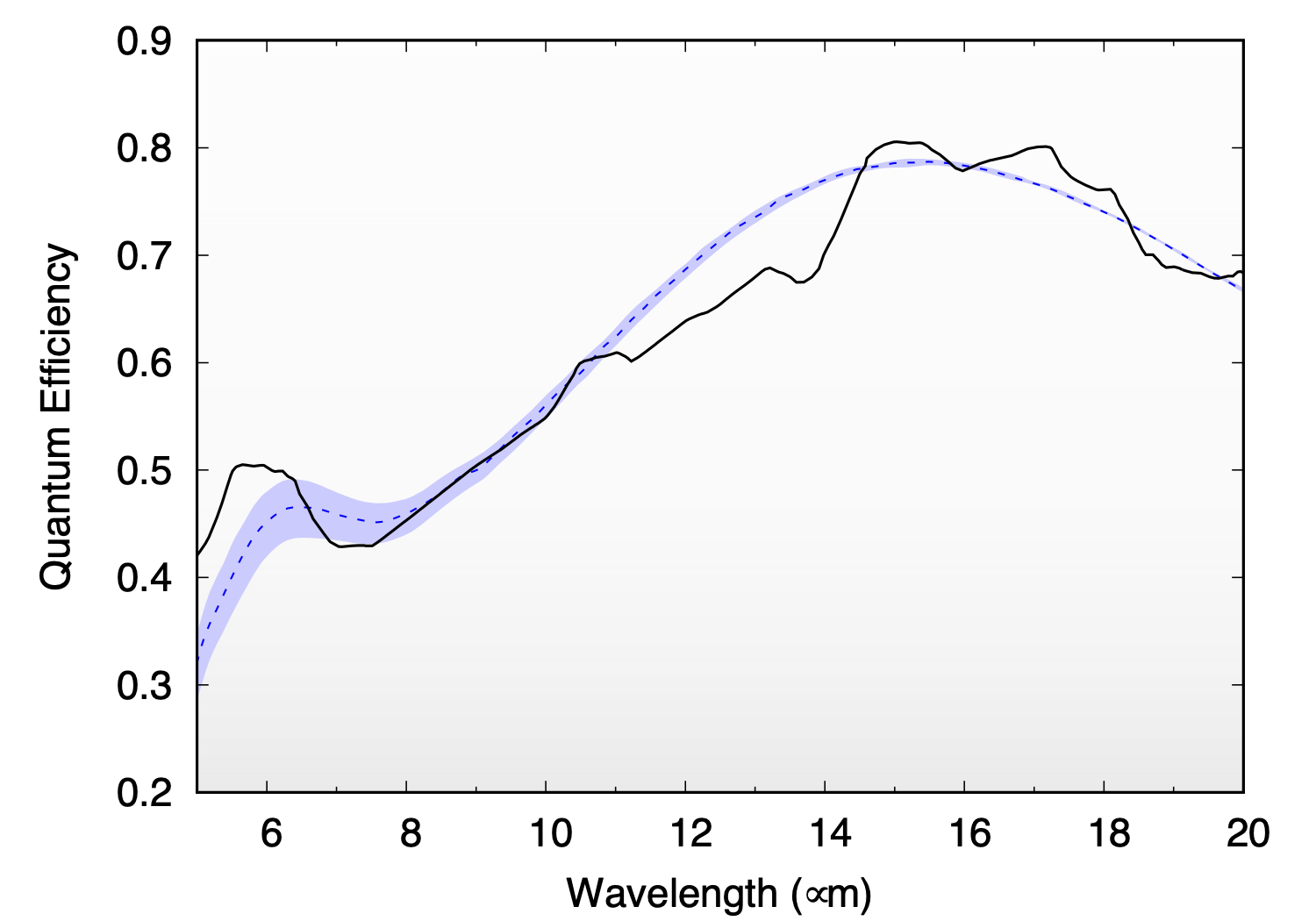}
\caption{Measured wavelength-dependent quantum efficiency for the MIRI baseline detector array modified to 
include the  calculated enhancement with the single-layer AR coating (solid black line, 
from \citet{ressler08}) compared with the  prediction from our model (dashed blue line). The shaded area shows 
the errors of the model due to the uncertainties in absorption coefficients, taken to be $\pm 6\%$ at 
10 $\mu$m and $\pm 12\%$ at 5 $\mu$m. }
\label{fig:QEt}
\end{figure}

A model was built combining these various effects. The absorption near 16 $\mu$m in the transparent contact 
was normalized to match the measured QEs.  The resulting dependence of QE on wavelength is shown in 
Figure \ref{fig:QEt}. The model provides a 
satisfactory fit to the measurements, and this fit is improved 
if the ``high'' values of the absorption coefficient are adopted at the shorter wavelengths, where high means 
those increased in accordance with the estimated errors in \citet{woods11}. In this model, the buried contact 
absorbs 18\% of the signal at 16 $\mu$m, qualitatively similar to the values obtained previously. 

The agreement of our model with the measurements of the detector array with a single-layer  anti-reflective 
coating optimized for 16 $\mu$m demonstrates that the internal operation of the detectors is well understood 
for the 5 $-$ 10 $\mu$m range. We therefore confirm that even higher QEs are expected in this range with a 
suitably optimized  antireflective coating. \citet{ressler08} show a predicted value of $\sim$ 60\%.

\subsection{Detector gain?}

\begin{figure*}[!th]
\includegraphics[width=0.98\textwidth]{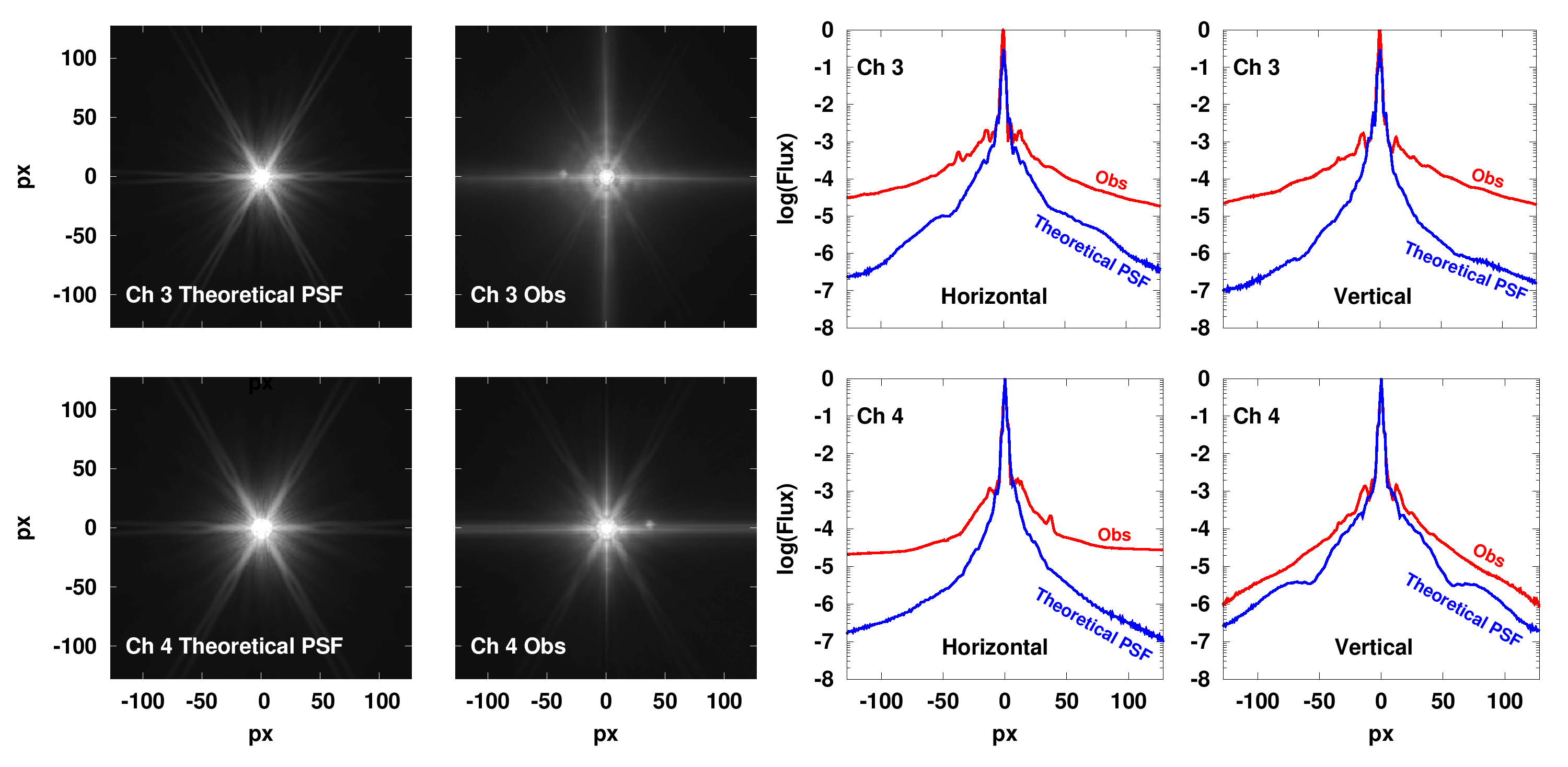}
\caption{The cross artifact (or ``banding'') is very apparent in the observed Ch 3 (5.7 $\mu$m) and 4 (7.8 $\mu$m) IRAC in-flight 
PSF images (downloaded from the Spitzer Science Center website). The artifact is not present in the STinyTim theoretical 
PSFs (John Krist), which includes the diffraction from the optical elements up to the focal plane of the instrument. The images
on the {\it extreme left} show the theoretical PSFs, the following columns the observed PSFs, while the 
plots on the {\it right} show the pixel values along the rows and columns centered on the source. The 
artifact contrast is higher in the Channel 3 data and not as apparent in the vertical columns at 
Channel 4 as they are for the horizontal rows due to a ``pullup'' effect \citep[see][for details]{hora04}.\label{fig:iraccross}}
\end{figure*}

The agreement between model and measurements in Figure \ref{fig:QEt} emerges in a straightforward way, based 
on a careful accounting of all the relevant effects. In an alternative approach, \citet{woods11} invoked 
detector gain at the wavelengths short of 10 $\mu$m, i.e. a quantum yield $>$ 1.\footnote{There are two 
common definitions of quantum yield (QY). In one, used by \citet{woods11}, it describes the number of charge 
carriers produced per incident photon. In the other, which we prefer, there are two terms: (1) the QE is the 
fraction of incoming photons absorbed to produce free charge carriers, i.e., a peak of 0.78 or 78\% in Figure 
\ref{fig:QEt}, and (2) the QY is the average number of charge carriers produced per absorbed photon, i.e. 
$\sim$ 1.66 at 5.6$\mu$m in the \citet{woods11} analysis. Under their  definition, the QY would be the QE 
multiplied by the QY as we define it. In our definition, the detector noise can be as small as the square 
root of the product of the photon flux and the QE, and QY $>$ 1 leads to extra noise terms. }  We find that this is not 
necessary and is in fact unlikely. QY $>$ 1 is improbable in extrinsic photoconductors (for photons below 
their bandgap energies) because the impurities that need to be ionized are so dilute in the crystal. In 
addition, gain in these detectors would be associated with increased noise above the simple square root 
of the number of absorbed photons.  Two effects are relevant: (1) the excess electrons over those for 
QY = 1 are not independent; the number of independent events is the number of absorbed photons, not the 
number of resulting photoelectrons; and (2) the statistical 
fluctuations in the number of photoelectrons produced per photon means that the noise will be increased 
further. As an example of the first effect, at 5.6 $\mu$m, Woods et al. (2011) predict QY = 1.66; 
therefore the number of independent absorbed photons will have been overestimated by this factor. Assuming 
the signal to noise ratio (SNR) goes as the square root of the number of absorbed photons, the SNR will 
be overestimated by a factor of 1.29. We simulated the second effect with a Monte Carlo calculation. This calculation allowed each simulated 
photon to produce 1, 2, or 3 photoelectrons and calculated the resulting signals as a 
1000 second integration sampled every 3 seconds (analogous to the operation in MIRI). The resulting integration ramp was fitted by 
linear regression and the results compared as a function of the QY.  For QY = 1.66, the prediction from this simulation is that 
the noise should be increased a further factor of
1.14. That is,  the value  of the QY found by Woods et al. (2011) implies that the  noise at 5.6 
$\mu$m should be elevated by a total factor of 1.29 $\times$ 1.14 = 1.47 over that estimated from 
the square root of the number of collected charge carriers.

\begin{figure*}[!th]
\center
\includegraphics[width=0.98\textwidth]{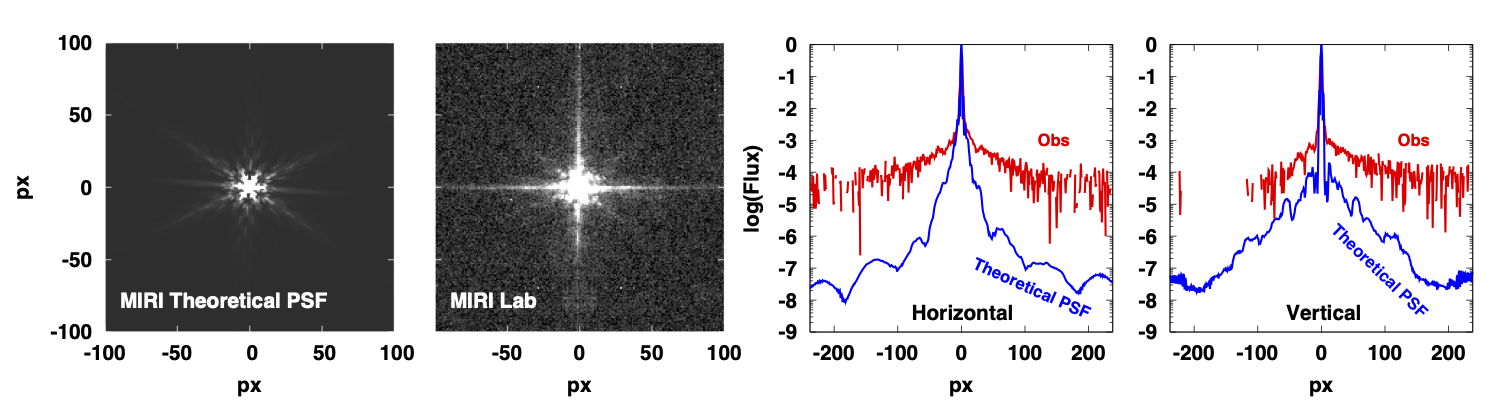}
\caption{
{\it Left panels:} The MIRI PSF generated with WebbPSF v.\ 0.8.0 \citep{perrin12,perrin14} at 5.6 
$\micron$ and a PSF mask image taken with the MIRI flight detector during CV2 (second cryo-vacuum). 
The cross artifact is easily identified in the observed images (additional artifacts related to the 
laboratory setup have been artificially removed from the image for clarity). The cross artifact can 
be traced up to at least $\pm$ 400 px in the high contrast laboratory image. {\it Right panels:} 
The row and column pixels along the cross in the laboratory and modeled PSFs. The diffractive pattern 
increases the PSF flux by 3 orders of magnitude along the cross, up until it disappears in the 
background noise.
\label{fig:miricross}}
\end{figure*}

This prediction has been compared with the measured noise from the MIRI detectors. We used calibration 
data obtained in the {\it JWST} Cryo-Vacuum Test 3 (CV3) to determine the noise. A series of measurements 
was taken in each imager filter (except 25.5 $\mu$m), with the total collected electrons adjusted to be 
approximately the same in all cases. These data were reduced to a series of images. Adjacent images were 
differenced to remove non-ideal behavior (e.g., reset anomaly, drifts) and photometry was conducted on the
resulting images using a pseudo-aperture approach in which the ``source'' region is square and the 
``sky'' is a surrounding square with the center removed, with all dimensions an integral number of 
pixels. The ``signal'' is then the total counts from the ``source'' minus the predicted number of counts 
adjusted to the same number of pixels from the ``sky''. 
This extraction unit was placed randomly (except it was always centered on a pixel) over the difference 
image to obtain a Monte Carlo set of noise measurements. This geometry allows very straightforward 
calculation of the expected noise since both source and sky contain an integral number of pixels.
We found a modest degradation of the detector noise behavior toward short wavelengths, by a factor 
of 1.15  at 5.6 $\mu$m. This effect could have a number of possible causes; for example, instability 
in the source temperature will modulate the output flux more at the short wavelengths than at 
the long ones.  Assuming the entire effect is true photon-noise-related, it is significantly less 
than predicted for the QY reported by Woods et al. (2011), i.e., the factor of 1.47 based on the photon statistics and our 
Monte Carlo simulation combined.

In the preceding section, we reproduced the full wavelength-dependent behavior of the MIRI detector 
quantum efficiency with QY = 1. Given this experimental evidence against values significantly above 
1, we believe QY = 1 is correct for these devices operating at wavelengths $\ge$ 5 $\mu$m.  

\section{``Banding'' or the ``cross'' artifact}\label{sec:thecross}

A general issue for back-illuminated semiconductor photodetectors arises when the absorption 
efficiencies are low, allowing some photons to penetrate to the frontside from which they are 
scattered back into the detector array. Particularly if the array is thick, they can be absorbed and detected far from their 
point of entry. For example, for ACS on {\it HST}: ``Long wavelength photons that pass through the 
CCD can also be scattered by the electrode structure on the [front] side of the device. This creates 
two spikes that extend roughly parallel to the x-axis. These spikes are seen at wavelengths longer 
than 9500 \r{A} in both the HRC and WFC'' \citep{ryon19}. Similarly for the STIS on {\it HST}: 
``Longward of 6000 \r{A}, the scattered light component becomes noticeable, primarily due to scatter 
in the STIS CCD and climbs to a level of $\sim 10^{-3}$ at 10,000 \r{A}'' \citep{gull2003}. 

A cross artifact due to this phenomenon at wavelengths $\le$ 10 $\micron$ with Si:As IBC detectors 
was first noticed in the Channel 3 and 4 IRAC/{\it Spitzer} images \citep{pipher04}; see Figure~\ref{fig:iraccross}. It can be 
described as a few rows and columns of bright pixels centered on a point source, with the artifact 
becoming brighter towards the source. The discussion below explores the causes of these internal 
scattering artifacts, focusing on Si:As IBC detectors, and  demonstrates a theoretical simulation 
of them. That is, while Section~\ref{sec:response} focused on the traditional type of response, 
trajectories (1) $-$ (4) in Figure~\ref{fig:arch}, we now focus on trajectory (5) and its further 
reflections. Since the analysis is computationally demanding, we simplify it by ignoring reflection
at the buried contact, which is only at a $\sim 4\%$ level. Enabling reflection would significantly 
increase computational complexity by allowing a large number of additional photon paths, which would need to be
tracked. Ignoring reflection at the buried contact has no significant effect on the results because 
it is an insignificant player in the response being modeled. We have verified this with a modification 
of our modeling code.

\begin{figure*}
\begin{center}
\includegraphics[width=0.98\textwidth]{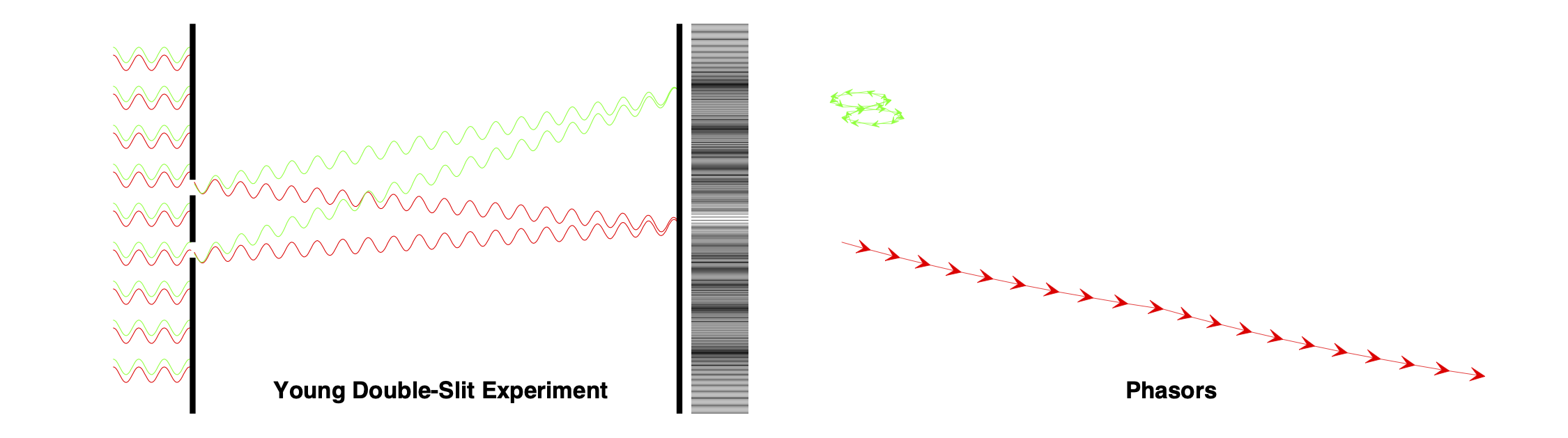}
\caption{The Young double-slit experiment explained with Quantum-Electrodynamics. Random phasors add up at
each imaging point and construct position dependent probability amplitudes. The square of the total probability
amplitude at each position defines their relative intensities. As the image shows, at the peak of the intensity, the individual
probability vectors add up to a large amplitude, while at the minimum they roughly cancel out and produce
a small amplitude.
\label{fig:youngqed}}
\end{center}
\end{figure*}

\subsection{Details of the Si:As IBC detector behavior}
\label{sec:detdetails}

 In Figure~\ref{fig:iraccross}, we show theoretical and in-flight observed point spread functions (PSFs) for the 
IRAC instrument, highlighting the considerable difference between the two. The Spitzer Science Center (SSC) pipeline notes that
the artifact exhibits flaring and that narrow-band images have a more complex artifact pattern. Since 
IRAC had a dedicated detector at each wavelength and the PSFs could be well characterized, there were 
no additional attempts to  understand the nature of the artifact theoretically.

A similar ``cross'' also appears in the MIRI detector images. In Figure \ref{fig:miricross}, we show the 
theoretical PSF calculated with WebbPSF v.\ 0.8.0 \citep{perrin12,perrin14} and images taken with the flight 
detector at the second cryo-vacuum test (with certain artifacts related to the laboratory setup artificially 
removed from the image), both at $5.6~\micron$. Remarkably, the artifact is present out to at least $\pm 400$ 
pixels in the high contrast laboratory images. As we show below, the artifact is a result of two optical 
properties of the detectors: 1) coherent light beams undergo diffraction off the metallic pixel contacts at 
the frontside of the detectors, and 2) inefficient photon absorption at shorter wavelengths ($\le 10$~\micron) in the 
IR-active layers and reflection off both the front and the back of the detector allows photons to travel 
through the detector multiple times. A pass through the detector and substrate three times (down, up, and down) 
produces an artifact extending to a maximum of $\sim 200$ px, meaning some of the $5.6~\micron$ 
photons cross the MIRI detector at least two times to be detected at a distance of over 400 px. 
At 400 px the WebbPSF contrast is $\sim 10^{-9}$, while the standard deviation of the background 
in the CV2 data is at a contrast of $\sim 5\times10^{-5}$. The cross is 1$\sigma$ above the 
background up to 150 px from the central source and is visually discernible to 400 px. 

\subsection{The Quantum Electrodynamic model of the detector}\label{sec:qed}

Since we were uncertain of details regarding the scattering/diffraction, we could not {\it a-priori} determine a simple 
specific model. Furthermore, our input light sources are more complicated than just a pinhole and realistic PSFs 
change as a function of wavelength. Therefore, we model the detector using the theory of Quantum Electrodynamics 
\citep[QED;][]{feynman48}; as an all-encompassing Monte Carlo-style (MC) method it allows us to model aspects of 
the system we otherwise may have ignored. 

\subsubsection{Basics of QED}

With QED modeling, all possible random directions of ray paths need to be explored, even if they
violate classical optical laws. Each ray path carries its own probability amplitude and phase angle 
(or complex phasor), where the phase angle advances with traveled distance as
\begin{equation}
    \Delta \varphi =  2\pi \frac{\Delta r}{\lambda}\;,
\end{equation}
where $\lambda$ is the wavelength of the tracer and $\Delta r$ is the traveled distance. As the photons 
enter the silicon substrate from the outer vacuum their wavelengths shorten by a factor of n$_{\rm sil}=3.41$,
i.e.\ a 5.6 $\mu$m photon will become a 1.64 $\mu$m photon. We quote the vacuum equivalent
wavelength of photons in our work; however, the calculations use the appropriately shortened wavelengths.
The phasors of unlikely paths cancel out, while the phasors of likely paths construct a large probability 
amplitude due to their similar phase angles. The resulting map of the square of the amplitudes yields the final 
relative probability of a photon being absorbed at a certain location and hence the intensity. In Figure 
\ref{fig:youngqed}, we explain the Young double-slit experiment with QED modeling, showing how the probability 
amplitudes at various locations add up. The total amplitude at the intensity peak is much larger than at the minima.

\subsubsection{Tracer paths}
\label{tracerpaths}

\begin{figure}
\begin{center}
\includegraphics[width=0.47\textwidth]{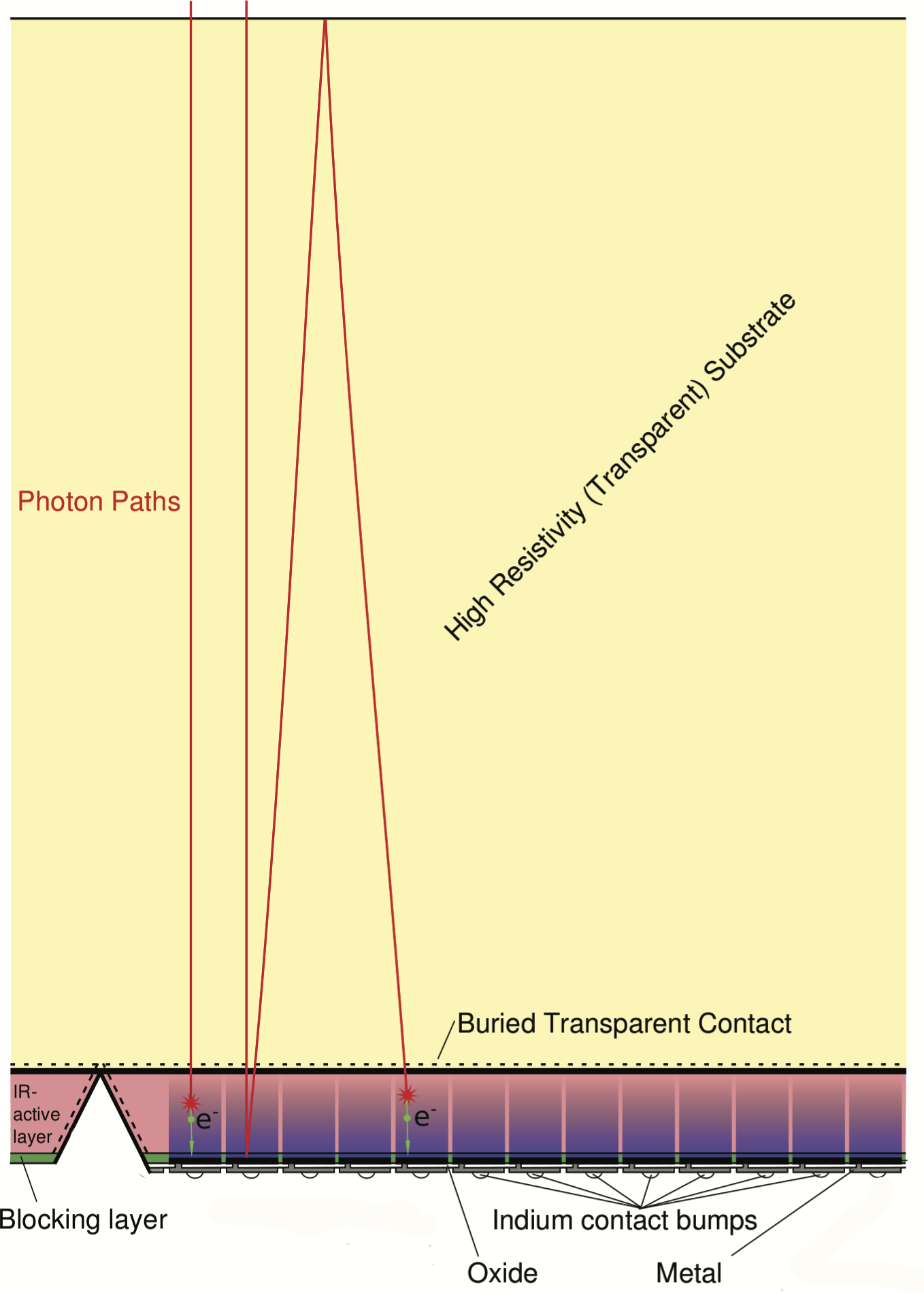}
\caption{To-scale cross section of a detector array, illustrating how photons on trajectories similar to 
(5) in Figure~\ref{fig:arch} can be returned to  array pixels far from their original entry points by 
total internal reflection at the array backside. The large thickness of the transparent substrate (relative to
the IR-active layer and pixel sizes) enables photons diffracted at even smaller angles to be later absorbed at 
pixels offset from their entry point.
\label{fig:detector}}
\end{center}
\end{figure}

\begin{figure}
\begin{center}
\includegraphics[width=0.47\textwidth]{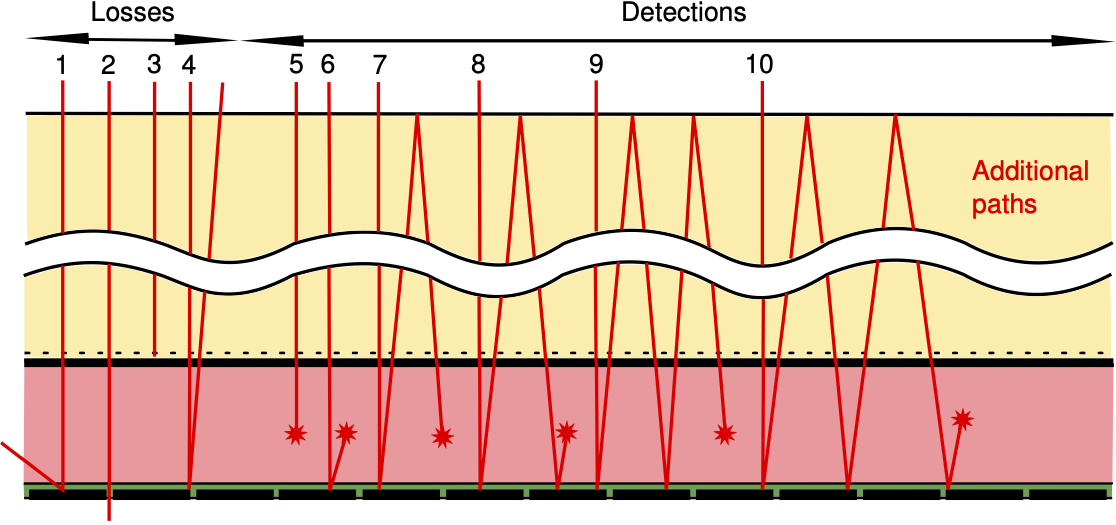}
\caption{10 most probable paths for a photon in the detector.
\label{fig:paths}}
\end{center}
\end{figure}

We first describe how photons creating the cross artifact travel in the detector and then how the QED model describes the
probability of each path. In Figure \ref{fig:arch}, we showed two possible paths with minor variations. In Figure 
\ref{fig:detector}, we show how diffracted photons ((5) in Figure~\ref{fig:arch}) can be returned to 
the active pixels by total internal reflection off the array backside.  We expand on this figure and show the most likely 
10 paths a photon may take in Figure~\ref{fig:paths}, with the probabilities of each path noted in Table \ref{tab:paths}. 
Paths 1 $-$ 6 are variations of those from Figure~\ref{fig:arch}, but paths 7 $-$ 10 illustrate the cases for photons 
trapped by total internal reflection and show how they may travel many pixel widths within the detector array before 
being absorbed, detected, or lost. This process is relatively efficient because of the good transparency of the buried 
contact at short wavelengths as shown in Figure~\ref{fig:trans}. Although theoretically there are an infinite number of 
possible paths for the photons to take through the detector, 99\% of the photons will travel on the most probable 
dozen or so path-types (see Table 
\ref{tab:paths}).

\begin{deluxetable}{llr}[t!]
\tablecaption{Various path types in the detector at 5.6 $\micron$; nomenclature based on Figure \ref{fig:paths}. 
The statistics is based on our best estimates for detector parameters; there will be deviations from these statistics depending on their exact values.\label{tab:paths}}
\tablecolumns{3}
\tablewidth{0pt}
\tablehead{
\colhead{Path ID} &
\colhead{Description} &
\colhead{Percent}
}
\startdata
\multicolumn{3}{c}{Losses}\\
\hline
1 & Exit at the side of the detector & 0.11\%\\
2 & Exit in the pixel gap & 9.39\%\\
3 & Absorption at the buried contact & 1.03\%\\
4 & Exit at the top of the detector & 1.79\%\\
\hline
\multicolumn{3}{c}{Detections}\\
\hline
5 & Absorption in 1$^{\rm st}$ pass & 27.09\%\\
6 & Absorption in 2$^{\rm nd}$ pass & 35.63\%\\
7 & Absorption in 3$^{\rm rd}$ pass & 11.91\%\\
8 & Absorption in 4$^{\rm th}$ pass & 7.68\%\\
9 & Absorption in 5$^{\rm th}$ pass & 2.56\%\\
10 & Absorption in 6$^{\rm th}$ pass & 1.65\%
\enddata
\tablecomments{The sum of all losses and the eleven most probable detected tracer paths 
encompass 99.98\% of all possible paths. Tracer path IDs larger than 10 are not shown in 
Figure \ref{fig:paths}; they are simply further crosses across the detector. Surprisingly, 
$\sim 11\%$ of the photons/tracers are lost through various modes. The first two detection 
paths will remain close to the image core and contain $\sim$63\% of the total flux (the detection
within the second passing - path \#6 - is higher than in the first, due to increased pathlengths
from off-axis diffraction). Around $\sim$26\% of the total flux will be spread out (i.e.\ not 
absorbed in the first two crossings of the IR active layer), remaining in the detector and 
raising the ``background'' signal or contributing to the cross artifact. Our model ignores an 
$\sim 4\%$ reflection at the buried contact.}
\end{deluxetable}

\begin{figure}
\begin{center}
\includegraphics[width=0.47\textwidth]{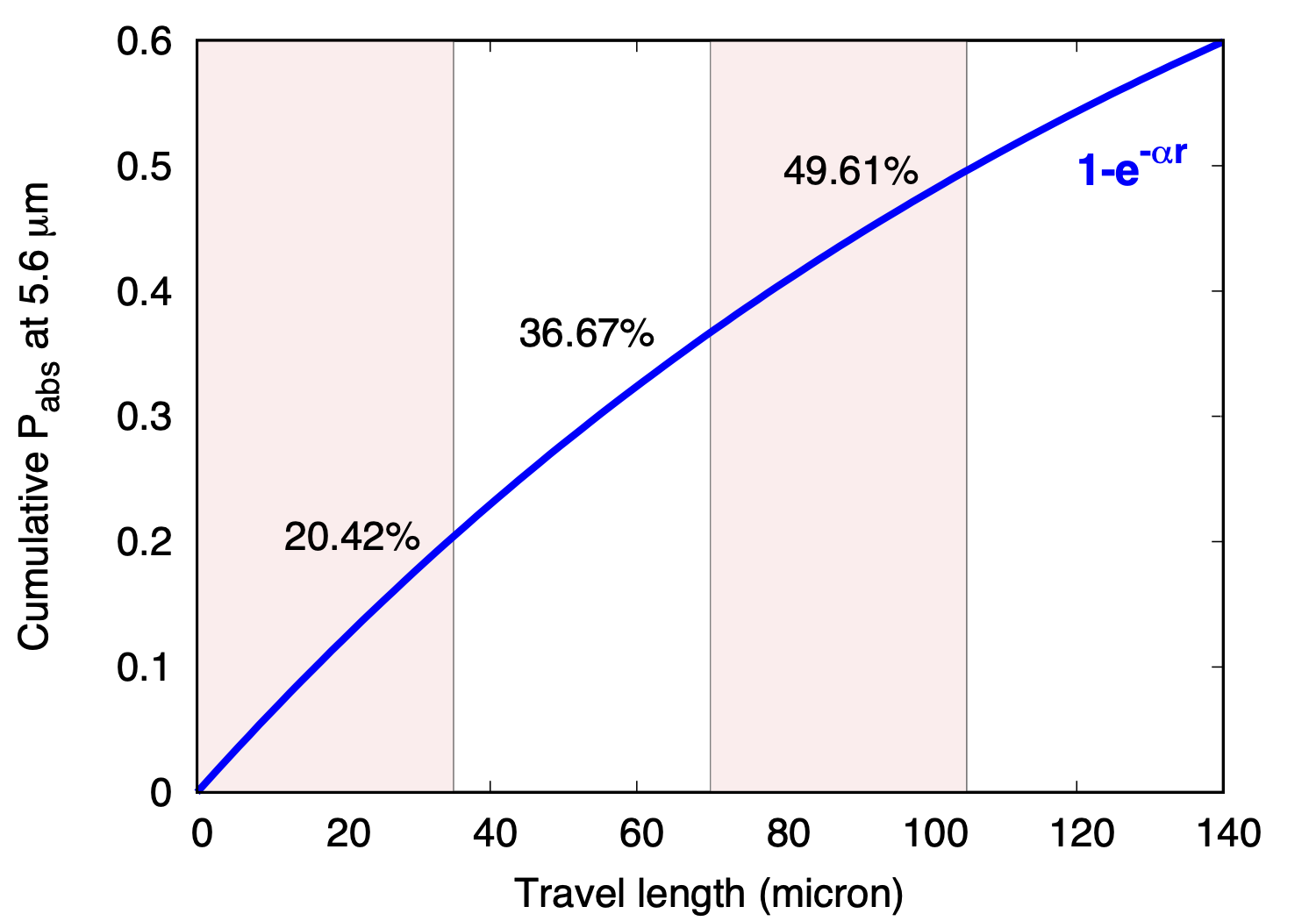}
\caption{The cumulative probability of a 5.6~\micron~photon being absorbed in an Si:As IR-active layer as a 
function of travel length. The layer depth (35~\micron) of the MIRI detector is shown and the probabilities of 
absorption marked for each full crossing of the IR-active layer. As the plot shows, even 2 full crossings does 
not guarantee a 50\% absorption probability for a 5.6~\micron~photon (this statistic obviously does not take 
into account the fact that the diffraction pattern will force photons to travel diagonally across the active 
layer, thereby increasing the travel lengths and their absorption probabilities for each crossing). However, 
a 17.8~\micron~ photon will have an over 90\% probability of absorption in a single crossing.
\label{fig:abs}}
\end{center}
\end{figure}

When describing the QED model, we use the word ``tracer'' to describe a particular photon 
path that is explored. We do this to differentiate it from paths that photons are more likely 
to take. Each tracer enters the detector with the same initial phase angle and unit amplitude.
The total path length covered by the tracers is cumulatively monitored and used to calculate the
phase angle when and if they are absorbed. As a tracer travels trough the buried contact, it may 
either get absorbed (and vanish from the calculation) or transfer through it, according to the 
absorptance illustrated in Figure~\ref{fig:trans}. As noted previously,
we ignore reflection at the buried contact, due to its low probability and the computational simplifications
this allows. The tracer then travels through the IR-active layer and either gets absorbed or survives 
its passage. The probability of absorption is described as
\begin{equation}
    P_{\rm abs} = 1.0-{\rm exp}(-\alpha {\rm r})\;,
\end{equation}
where $r$ is the distance traveled in the IR-active layer and $\alpha$ is the absorption efficiency, which
we derive from the fitted QE, i.e. the values as shown in Figure~\ref{fig:xsec} adjusted to optimize 
the fit as in Figure \ref{fig:QEt}. In Figure \ref{fig:abs}, we show how the probability of
absorption increases for 5.6 $\mu$m wavelength photons with the cumulative distance traveled in the 
IR-active layer.

\begin{figure}[t!]
\begin{center}
\includegraphics[width=0.47\textwidth]{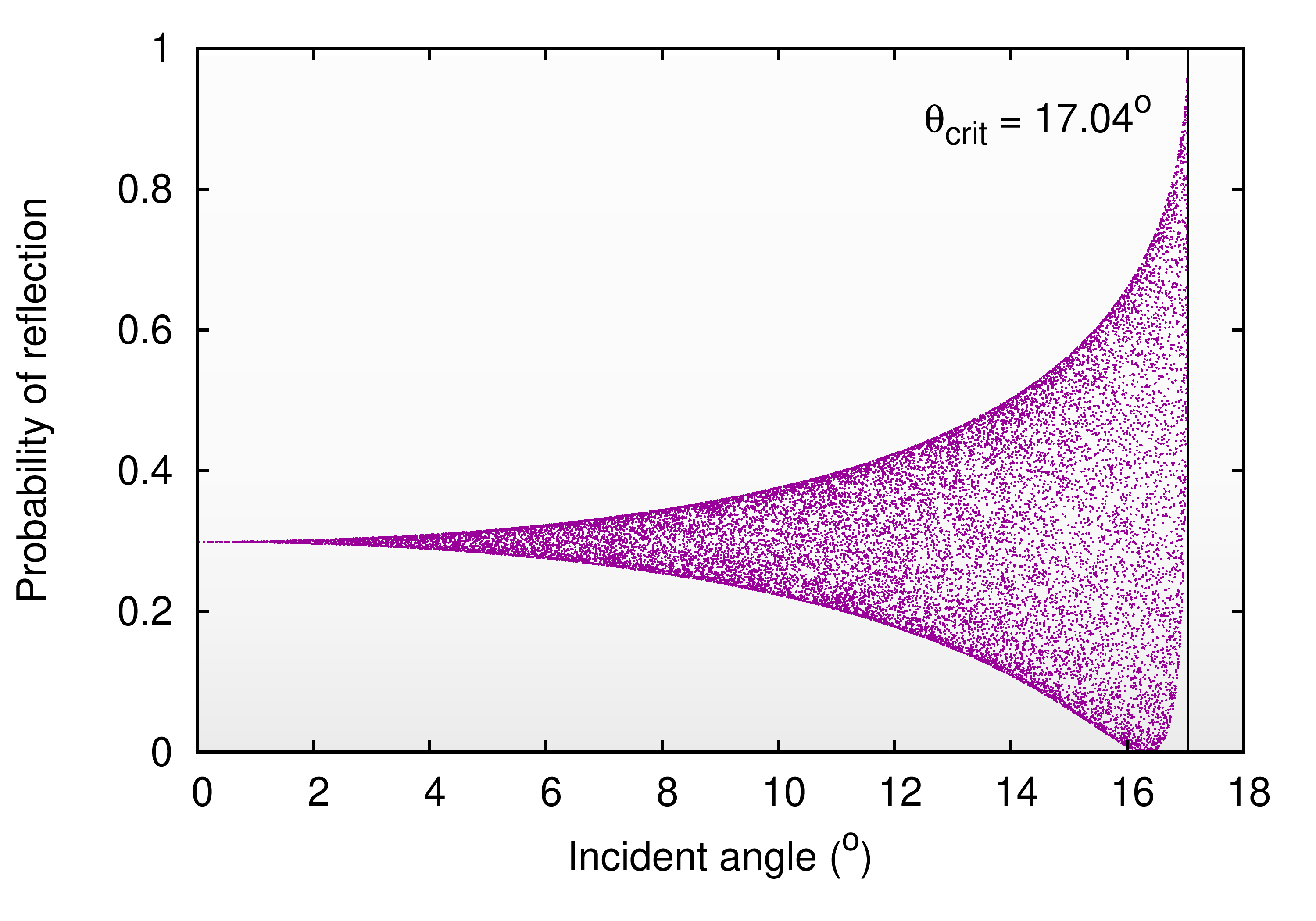}
\caption{Reflection probabilities at the top (``backside'') of the detector for 50,000 tracers returning 
back after reflecting off the metallic pixel contacts. At incidence angles larger than the critical 
angle of total internal reflection all tracers will remain within the detector wafer, however, 
tracers with smaller incidence angles will either reflect or exit the detector, depending on their 
random polarization angle. If an astronomical source is emitting polarized light, there will be 
offsets from our general prediction.
\label{fig:refl}}
\end{center}
\end{figure}

The total length that is travelled through the IR-active layer will depend on the incidence angle of the 
tracer, which is practically 0$^{\circ}$ for the first passage (ignoring minor variations due to OTA diffraction). 
If a tracer survives, it will get reflected off the metallic contacts at a random angle or disappear in the 
pixel gaps between the metallic contacts.
Since in this section we are modeling the diffraction pattern, and the only surface yielding one is the pixel lattice, 
we generate new random directions for the tracers only at the metallic contacts and evolve them 
following regular reflection laws at the  reflecting surface (hence the unusual reflection 
angle for the paths in Figures \ref{fig:arch} and \ref{fig:paths}). This allows us to calculate 
the model with a ``reasonable'' number of photons. A reflected/diffracted tracer will assume its new 
direction and continue to travel toward the detector backside, with similar physical considerations 
as on the way from the backside to the frontside. For incidence angles larger than 0$^{\circ}$, the 
travelled length will increase and therefore so will the absorption probability (see Figure \ref{fig:abs}) 
in the IR-active layer. The final fork in the road traveled by a tracer is at the backside, where it will either 
exit the detector or be reflected and repeat the previous cycle. Whether a tracer reflects or 
not depends on its incidence and polarization phase angles, and the probability is given by the 
Fresnel equations. If the incidence angle is larger than the critical angle of total internal 
reflection
\begin{equation}
    \Theta_{\rm crit} = {\rm asin}\left(\frac{1}{n_{\rm sil}}\right) = 17.04^{\circ}\;,
\end{equation}
then the tracer reflects with a nearly 100\% probability (since the AR coating is virtually lossless). 
If this incidence angle is less, then we need to 
apply the Fresnel equations. Since most astronomical objects are not polarized, the model picks a 
random polarization angle and calculates a reflection probability. In Figure \ref{fig:refl}, we 
show the reflection probabilities of 50,000 tracers in the code, highlighting the large range of 
probabilities considered for the majority of incidence angles.

Since the patterns will likely form on spatial scales proportional to the input wavelength (which is 
reduced by a factor of n$_{\rm sil}$ once the photon enters the silicon detector), we divide each pixel into 
sub-pixels and calculate the phasor amplitudes as if the photons were absorbed in the center of the 
sub-pixels. The default setting of the model is to use 1 \micron\ sub-pixels, which is close to 
Nyquist sampling the 1.64 \micron\ wavelength of a 5.6 \micron\ input photon after it enters the silicon. The model can also divide 
the IR-active layer into sub-layers, although this division has little effect. Absorption is 
calculated in the center of the sub-layers; if a single absorptive layer is assumed (as in the 
default setting), then absorption is projected to occur in the middle of the IR-active layer.

\subsubsection{Image flux scaling}

We sample tracers from the square root of the initial intensity distribution and track the position and phase 
of their phasors as they travel in random directions. Sampling from the square root of the intensity distribution 
is necessary, as photon probabilities (i.e.\ intensities) are defined as the square of the total phasor amplitude. 
Therefore, to select unit amplitude tracers with correct probabilities, we need to sample them from the square 
root of the input intensity distribution. If a tracer is absorbed, we add its phasor to the total phasor in the 
respective sub-pixel it was absorbed in. When choosing from an input image or Airy ring, spatial positions are 
determined randomly based on cumulative flux distributions, thereby selecting brighter input positions more 
frequently than fainter ones.

We also need to track the number of tracers absorbed and initiated, as it is necessary to scale 
the relative probabilities. The vast majority of tracer paths are low probability, but their 
low probability needs to be established with a large number of test paths. Therefore, the model records 
each probable photon path type separately, being able to distinguish between the most probable photon 
paths. This is important because the final relative intensity map is calculated as the square of the 
phasor amplitudes, scaled by the total number of photons absorbed relative to the number of photons the 
model used. In equation form, the total intensity of a particular path at pixel position $(x,y)$ would be described by
\begin{equation}
    I_{\rm i}(x,y) = \frac{N_i}{N_{\rm tot}} \times \frac{A^2_i(x,y)}{\sum_j A_j^2}\;,
\end{equation}
where $N_i$ gives the total number of tracers absorbed in path $i$, $N_{\rm tot}$ gives the total number of test tracers
used by the program, $A^2_i(x,y)$ the amplitude squared in pixel (x,y) for path $i$, while $\sum_j A^2_j$ is the total 
intensity captured with path $i$.

As many more photons are absorbed in the first photon passing, and probabilities are calculated as relative 
differences between sums, it would receive a significantly higher flux value if we did not correct for the 
number of photons each path (and layer) possibility (each with its own diffraction pattern) absorbed. 
This is an important aspect of QED; it calculates relative probabilities using sums. Furthermore, each 
absorption event is independent; there is no “interference” in QED in the classical sense (even though 
the final product is equivalent). Once a large enough number of tracers has been recorded for each possible 
path and absorption layer, we calculate intensity maps for each path-type and sum them up per pixel. Finally, 
the intensities of all path-types are summed for the total intensity per pixel.

\begin{figure*}[t!]
\begin{center}
\includegraphics[width=0.98\textwidth]{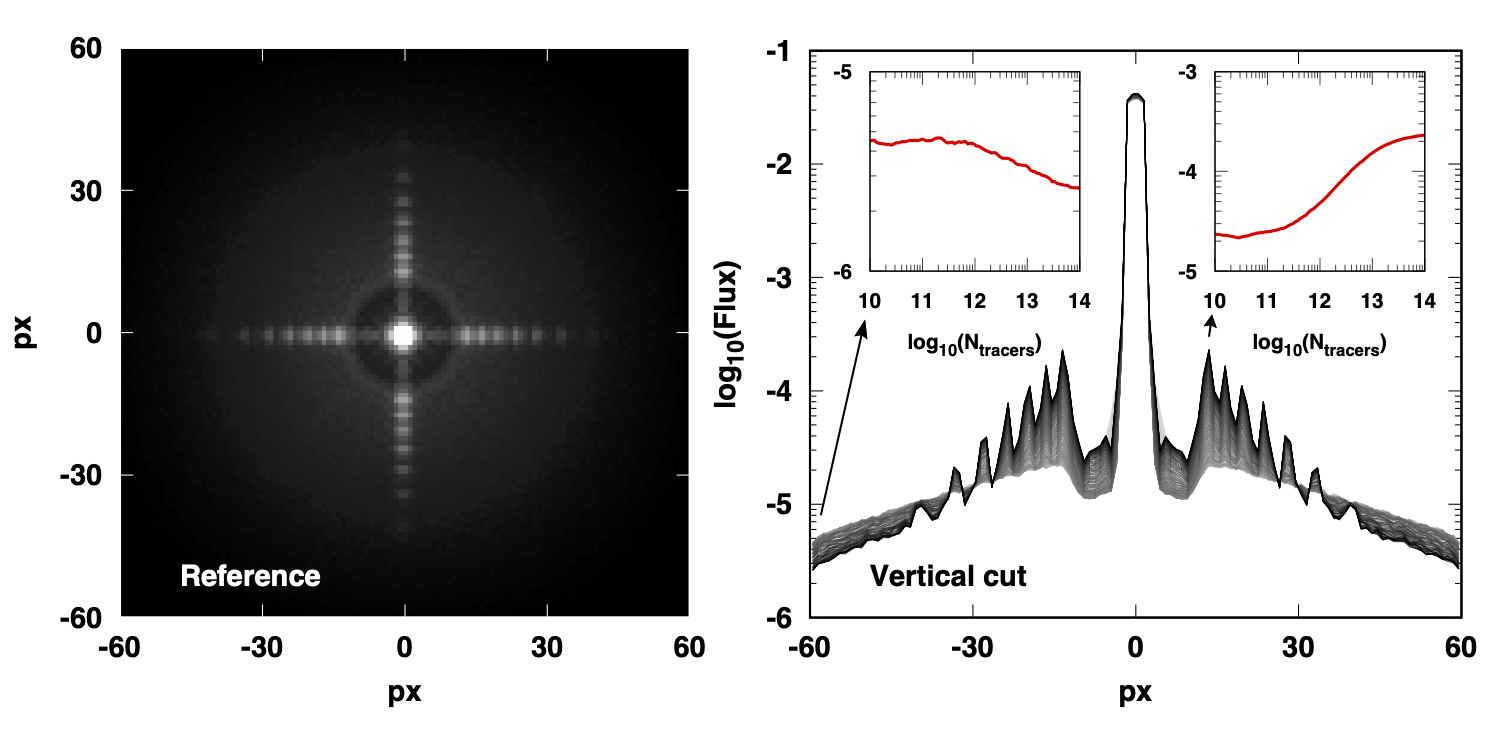}
\caption{{\it Left panel:} Reference model GimMIRI image using a 50 \micron\ radius pinhole as input flux 
distribution at an input wavelength of 5.5 \micron, modeled with $10^{14}$ tracer phasors. The variables 
of the Reference model are listed in Table \ref{tab:refs}. The cross artifact is apparent as well as a large halo. 
{\it Right panel:} The pixel values in the Reference model across a vertical cut centered on the pinhole. The flux 
values are shown for models between $10^{10}$ and $10^{14}$ tracers, with the lines getting darker for larger model 
tracer values. The convergence of the flux distribution for pixels (at rows -60 and 14) is shown in the sub-panels. 
A concrete criteria for convergence was not set; images were examined on a case-by-case basis. For the reference model,
the two example pixels shown have levelled off to a similar relative variation of $10^{-16}$/tracer. With two differently
illuminated regions (offset by an order of magnitude in flux) fluctuating around a similar low level of variation,
the reference model was determined to have mostly converged at $10^{14}$ tracers, which took 59 hours of calculation time 
on two Nvidia TITAN BLACK GPUs.
\label{fig:ref}}
\end{center}
\end{figure*}

\subsubsection{Our modeling code GimMIRI}
\label{sec:code}

Unfortunately, MC algorithms are notoriously slow to converge, therefore our modeling code 
utilizes graphical processing units (GPUs). GPUs, with their parallel processing capabilities, provide
an ideal environment for calculating the trajectories of independent tracers; hence they are widely used in
ray tracing applications. We opted to encode our model using the Nvidia CUDA language, since the 
University of Arizona has a substantial Nvidia GPU cluster server\footnote{\url{https://elgato.arizona.edu/}} 
and we have also received an academic hardware donation directly from Nvidia. The code we developed, called
GimMIRI (GPU image for MIRI), is a multi-GPU single-node program written in C/CUDA, allowing it to use as many 
GPUs as are available (or requested) on a single computer node. \footnote{Download from \url{https://github.com/merope82/GimMIRI}} 
The model only has a few basic input variables and a multitude of tweakable detector parameters that are defined and 
hard-coded in the source. The calculations are done at a single wavelength (which is one of the input variables) 
for each execution of the code and therefore  if an extended bandpass is studied, multiple models need to be run 
and their results co-added. The user can select between three flux input distribution types: 1) any fits file of 
their liking, 2) a pinhole of chosen radius, or 3) an Airy function (specified by the observing aperture). 
If a WebbPSF \citep{perrin14} PSF fits file is used as input, the code sets the wavelength and oversampling 
rate for the calculation from the image header (which can be overwritten). Other than output frequency and naming 
options, the only variable the user sets for the compiled code is x and y pixel offsets.

Each individual CUDA thread calculates the path of a single tracer, generating numerous independent random numbers during
the calculations: assigning new random angles of propagation at the metallic contacts, deciding whether a tracer is absorbed
in the IR-active layer or the buried contact, and setting reflection probabilities based on random polarization 
angles at the backside of the detector. 
To ensure that each thread generates independent random numbers, all threads on all cards have their unique random generator 
seed that is stored on the graphical cards. Up until absorption, all threads calculate independently, and upon absorption an
atomic addition process (serialized) adds the tracer phasor values to the specific sub-pixel's cumulative value. Since it is not likely
that two threads will add to the same exact subpixel, the atomic addition does not slow down the code. At requested
intervals the code writes a new fits image, with the image header containing all relevant information.

\subsubsection{Variables of the model}

The variables of the model define the detector architecture, the finesse of the calculation, and 
the size of the output image. In Table \ref{tab:refs}, we define these variables and their values used in
our reference model. Since the exact values of some of these variables are not available, we use reasonable 
assumptions for the MIRI detector. We execute a Reference model at 5.5 $\micron$,
assuming a 50 $\micron$ radius pinhole as input flux distribution. While the pinhole is obviously not 
representative of an actual PSF, its simplicity and concentrated light allows our model to converge faster 
and enables a clearer study of each detector parameter. In Figure \ref{fig:ref}, we show our Reference model image.

\subsection{Exploring the parameter space}\label{sec:variables}

\begin{deluxetable}{llr}
\tablecaption{Variables used in the Reference model.\label{tab:refs}}
\tablecolumns{3}
\tablewidth{0pt}
\tablehead{
\colhead{Variable} &
\colhead{Description} &
\colhead{Value}
}
\startdata
$D_{\rm act}$      & Width of IR-active layer          & 35.0  \micron \\
$D_{\rm top}$      & Total height of detector          & 500.0 \micron \\
$s_{\rm gap}$      & Pixel gap size                    & 2.0   \micron \\
$s_{\rm px}$       & Pixel pitch                       & 25.0  \micron \\
$s_{\rm sub}$      & Sub-pixel size                    & 1.0   \micron \\
$\lambda$          & Input wavelength                  & 5.5   \micron \\
$r_{\rm pin}$      & Pinhole radius                    & 50    \micron \\
$N_{\rm path}$     & Number of paths considered        & 9             \\
$N_{\rm lay}$      & Number or IR sub-layers           & 1             \\
$n_{\rm sil}$      & Refractive index of material      & 3.4127        \\
$N_{\rm x}$        & Output image size in x            & 120           \\
$N_{\rm y}$        & Output image size in y            & 120           \\
$\Theta_{\rm off}$ & Offset in incident angle (normal) & 0.045$^{\circ}$ \\
$\Gamma_{\rm off}$ & Offset in incident angle (CCW)    & -13.75$^{\circ}$
\enddata
\end{deluxetable}

\begin{figure}[t!]
\begin{center}
\includegraphics[width=0.47\textwidth]{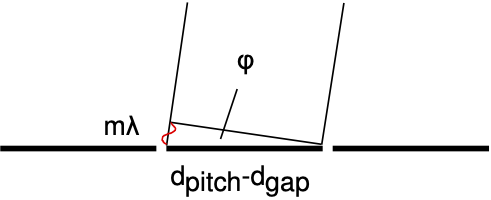}
\caption{The classic wave interference explanation for diffraction off a periodic lattice.
\label{fig:classic}}
\end{center}
\end{figure}

The diffraction pattern that is imaged depends on the various geometric parameters of the detector and the
observed wavelength. As a first order approximation, the classic wave interference explanation of diffraction
can help place each of them in context. In Figure \ref{fig:classic}, we display the direction of two beams 
at the top of a pixel-defining metallic contact, where the two beams constructively interfere at the $m=1$
order. Based on this interpretation, one would expect constructive interference, where 
\begin{equation}
    m \lambda = \left(d_{\rm pitch}-d_{\rm gap}\right) {\rm sin} \varphi\;.
\end{equation}
In the far field extreme, the two beams meet at infinity and the pixel offset can be calculated as
\begin{equation}
    \Delta P = 2 \frac{D_{\rm top}}{s_{\rm px}} {\rm tan} \varphi\;.
\end{equation}
Since the height of the detector ($D_{\rm top}$) is only 20 times that of the lattice length (and around 300 times
the wavelength), the far field approximation is not necessarily adequate for modeling, but we can use it for demonstration
purposes. For the reference model, the pixel offsets of the $m=1$ through 10 order peaks are at 2.8, 5.7, 
8.6, 11.7, 15.0, 18.5, 22.5, 27.1, 32.5, and 39.3 pixels, according to the far field approximation. In fact, we 
see peaks in the diffraction pattern for the reference model at $\Delta P=13.4$, 16.5, 19.6, 23.5, 27.9, 33.0, and 
39.6 pixel offsets, corresponding to the $m=4$ to 10 orders. The $m=1$, 2, 3 orders fall inside the critical angle of 
total internal reflection and are therefore not produced at a high contrast level. The agreement between the far field
approximation and model results is worse for the lower orders, as the pathlength of the rays is shorter for them. 
For $m\ge6$, the difference is less than a pixel for the reference model. Obviously, the level of agreement will depend on the 
detector architecture and the wavelength of observation. At $m=15$, the diffraction angle would be larger than 
90$^{\circ}$ and therefore it is not physically plausible; the $m=14$ diffraction angle is 78.8$^{\circ}$ and 
the pixel offset is $\Delta P = 202.2$ px. This calculation also shows that any artifact that extends over this 
distance must be produced by more than a single reflection off the metallic contacts.

\begin{figure*}[t!]
\begin{center}
\includegraphics[width=1.0\textwidth]{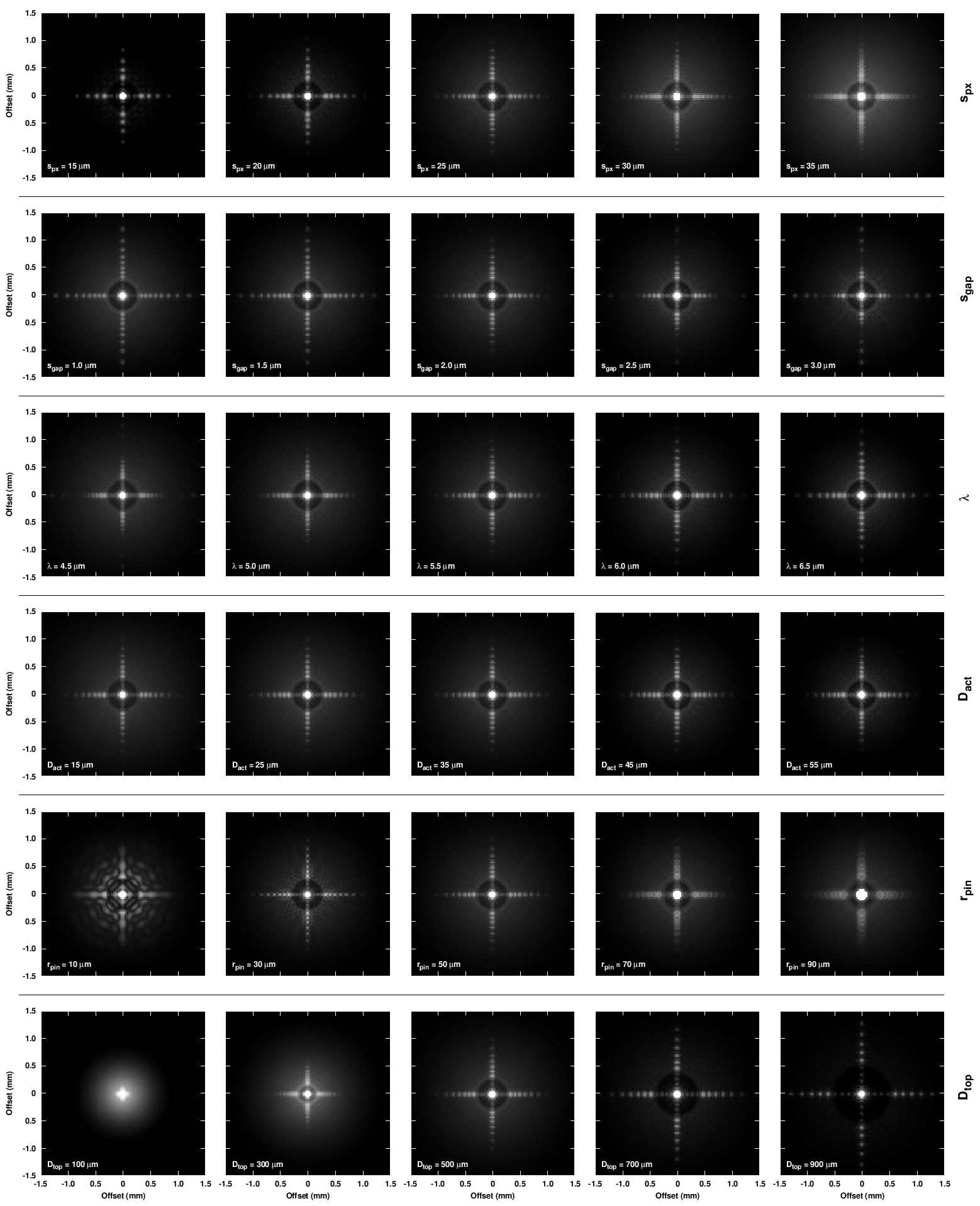}
\caption{The diffraction pattern as a function of detector variables modeled with GimMIRI. The central column 
is the reference model. The variables are summarized in Table \ref{tab:refs}. 
\label{fig:variables}}
\end{center}
\end{figure*}

We modeled the diffraction pattern as a function of varying values for the pixel pitch ($s_{\rm px}$), the pixel
gap ($s_{\rm gap}$), the input wavelength ($\lambda$), the detector substrate height ($D_{\rm top}$), the size
of the pinhole radius ($r_{\rm pin}$), and the width of the IR-active layer ($D_{\rm act}$). In Figure \ref{fig:variables},
we show the GimMIRI image as a function of these variables. The details of the pattern change, but the overall pattern 
remains and is dominated by the diffraction off the square grid of contacts. Other dominant patterns would result 
from other contact geometries.

We conducted a number of tests to be sure of the validity of our models. We checked for numerical
convergence in Figure \ref{fig:ref}. The GimMIRI reference model is converged by 10$^{14}$ tracer phasors.
When modeling realistic PSFs, more tracers are necessary, therefore we will check for convergence again when modeling
the {\it JWST} PSF responses at higher fidelity. We also checked dependence on the sub-pixeling and the number of sub IR-active 
layers we model. Nyquist sampling with sub-pixeling is critical, however, sub-layering the IR-active layer
is not as important, as long as absorbed photons are projected to be absorbed in the center of the layer.

\section{Verifying our QED model}

We verify our QED diffraction model against the artifact pattern observed in the MIRI CV2 tests. For this test,
we used the WebbPSF model of MIRI at 5.6 $\mu$m as the input flux distribution. It models the complete JWST 
optical path up to the instrument focal plane, and therefore provides the ideal input flux distribution
for our model. To simulate a realistic smooth input PSF, we used a 16 times ovsersampled WebbPSF model for our
calculations.  Calculating the GimMIRI image (to $10^{15}$ tracers) took a total of 1200 hours on two Nvidia 
TITAN BLACK GPUs.

\begin{figure*}[t!]
\begin{center}
\includegraphics[width=0.9\textwidth]{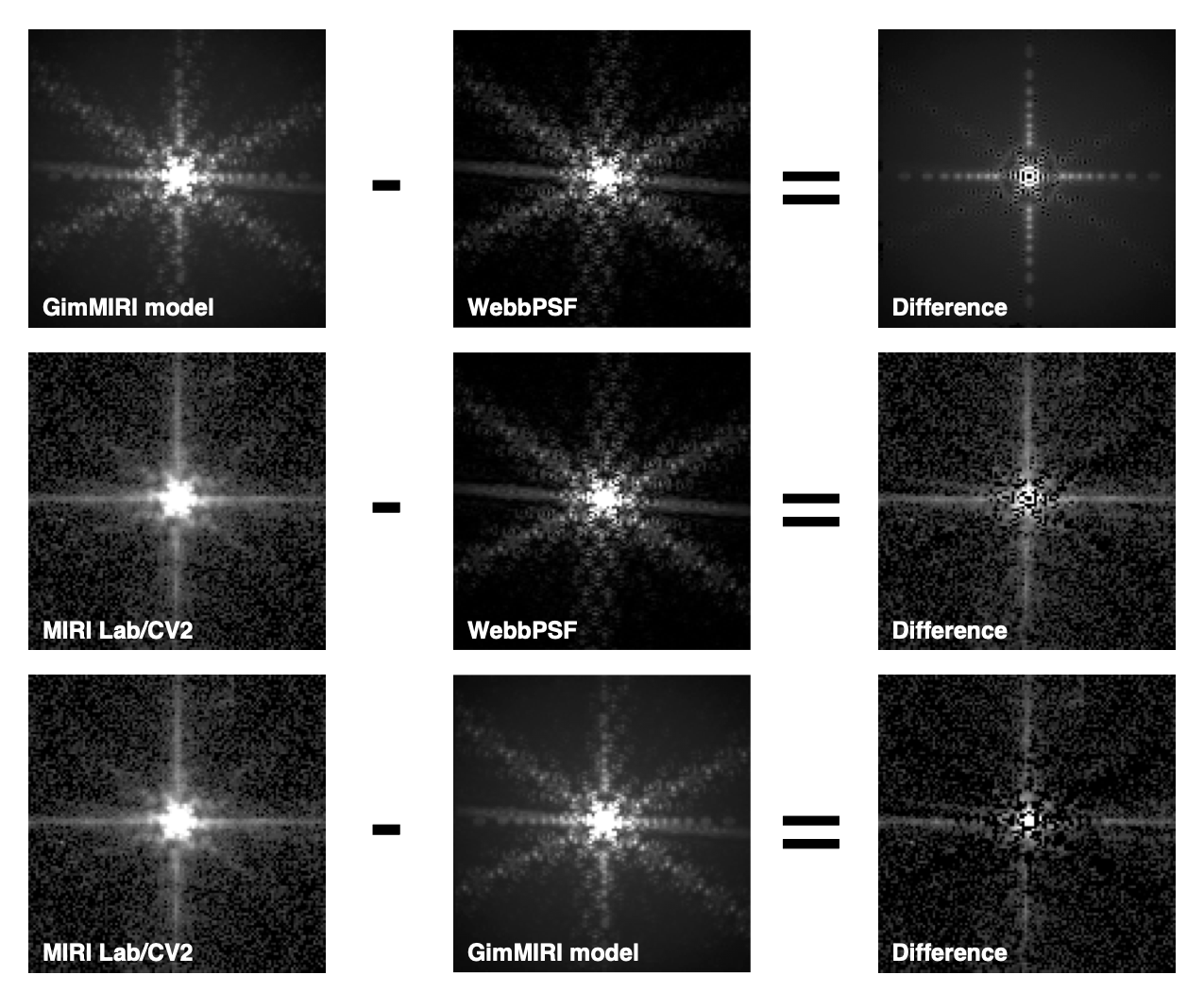}
\caption{{\it Top Row:} Comparing the GimMIRI and WebbPSF model images at 5.6 \micron. The difference image highlights
the location where the diffraction displaces $\sim$ 21\% of the incoming flux, mainly into the diffraction spikes and
a broad halo. {\it Middle Row:} Comparing the observed CV2 image and the WebbPSF model image. The difference image is 
generally similar to the one above, although there are some discrepancies for the inner region and the cross artifact 
is of greater extent. {\it Bottom Row:} Comparing the observed CV2 image and the GimMIRI model. The difference image 
shows that the model reproduces most of the artifacts in the observed image. However, it under-represents the extent 
of the cross artifact and has some low-level remaining discrepancies in the inner zone. These remaining issues can
be addressed by simulating the image for the full bandwidth of the filter rather than for a single wavelength and 
possibly by using more tracer particles or a model addressing just this region. All images are scaled linearly between
intensity levels of 0 and 0.0005 (where the total intensity of the theoretical PSF equals 1 and peaks at 0.15 in its 
core). See more details on this figure in the text.
\label{fig:refmiri}}
\end{center}
\end{figure*}

The WebbPSF model has an integrated unit intensity (by definition), while the modeled GimMIRI PSF has an integrated
intensity of 0.835. This shows that 83.5\% of the incoming flux remains within the detector; the increase compared 
with the QE calculated for the central response only in Section~\ref{sec:response} is an indication of the extent of 
the diffraction effects. The number is also slightly lower than the cumulative absorbed value given in Table
\ref{tab:paths}, as the verification model ``only'' encompassed a 200 px $\times$ 200 px area, resulting in higher
losses at the detector edge. The peak intensity of the PSFs, however, scale by a factor of 0.624, which agrees with
the QE values estimated by \cite{ressler08} and in Section~\ref{sec:response} for a device with an AR coating 
optimized for 6 $\mu$m. The difference between the two scalings highlight that a significant portion (21.2\%) of the
flux gets placed into both the diffraction spikes and into a halo surrounding the peaks, and also the problems posed 
by the artifacts when calibrating spectral data.

In Figure \ref{fig:refmiri}, we compare the theoretical {\it JWST}/MIRI PSF calculated with WebbPSF to the complete 
PSF calculated with GimMIRI, and with that measured using a flight hardware at the second cryo-vacuum tests (CV2).
All images are at 5.6 \micron\ (at native pixel scale).  The cross artifact is apparent in both the observed PSF 
and the modeled one, although modeling with additional photon tracers would likely strengthen its signal in the 
final model image. The top row of Figure \ref{fig:refmiri} shows the output GimMIRI PSF image and the WebbPSF model
that was used as the input for the modeling, as well as their difference. The top-right panel highlights where the
internal diffraction places the flux removed from the PSF core. The middle-row of Figure \ref{fig:refmiri} shows 
the difference between the measured CV2 PSF and the WebbPSF model. It is clear that the WebbPSF is not adequate 
to model the observed diffraction structure. The bottom row of Figure \ref{fig:refmiri} compares the CV2 image 
with the GimMIRI one. Our model removes the PSF core and most of the diffraction pattern, but not the complete 
structure. This is due at least in part to the fact that the laboratory image was taken using the F560W MIRI 
filter, which has a broad transmission window within 5.0 and 6.2 \micron, while the GimMIRI model is at a single 
wavelength. The broader transmission profile will result in a smooth diffraction spike, which the single wavelength model 
cannot replicate.

Our model reproduces the cross artifact produced by tracers up to path \#8 in Figure \ref{fig:paths}, which can
physically extend up to $\sim$ 200 px. The artifact can be traced further out at very low levels (see Section
\ref{sec:detdetails}) in actual images. While our model does include these unlikely tracer paths, modeling 
constructive interference with QED tracers at those distances is very expensive computationally. Further work 
will be necessary to enable quicker (and possibly less accurate) modeling of the diffraction structure. 
Regardless, GimMIRI provides a complete model of all the attributes of the detector physics, thereby providing 
a baseline for future approximations. 

\section{Summary}\label{sec:summary}

We have addressed two phenomena regarding the short wavelength (5 $-$ 10 $\mu$m) performance of Si:As IBC 
detector arrays: (1) quantum efficiency; and (2) image artifacts. Our study confirms:

\begin{itemize}

\item{Quantum efficiencies in the 5 $-$ 10 $\mu$m range of $\sim$ 60\% are attainable with these detectors 
when suitably anti-reflection coated.}

\item{This performance is achieved without quantum yields (QYs; i.e. the number of free charge carriers 
yielded per absorbed photon) $>$ 1. There should be no excess noise as would result if QY $>$ 1.}

\end{itemize}

At wavelengths of 5 $-$ 10 $\mu$m, these detectors show a large-scale cross-like imaging artifact.
Related behavior is seen in other back-illuminated solid-state detectors where the absorption 
efficiency is low, e.g. conventional CCDs in the near infrared, allowing photons to scatter off contacts 
and other structures on the detector frontside. We show that

\begin{itemize}

\item {The underlying cause of these artifacts is low absorption efficiency at the particular wavelength
and diffraction off inter-contact gaps}

\item{Photons can be diffracted to sufficiently large off-axis angles ($> 17^{\rm o}$ in silicon) 
that they are totally reflected when they reach the detector backside.}

\item{Consequently, these photons can be trapped in the detector/substrate wafer and can travel 
large distances until they are absorbed or escape, often leading to their detection far from their 
point of entry.}
\end{itemize}

\acknowledgments

This research made use of Tiny Tim/Spitzer, developed by John Krist for the Spitzer 
Science Center. The Center is managed by the California Institute of Technology under 
a contract with NASA. This material is based upon work supported by the National 
Science Foundation under Grant No. 1228509. We are grateful for the generous 
hardware donation from the Nvidia Corporation.

\vspace{5mm}

\bibliography{ms}

\end{document}